\documentclass[a4paper,amsmath,amssymb,pre,twocolumn,superscriptaddress]{revtex4-1}

\usepackage{hyperref}
\usepackage{eepic}
\usepackage{graphicx}
\usepackage{overpic}
\usepackage{subfigure}
\usepackage{latexsym}
\usepackage{bm}
\usepackage{amsmath}
\usepackage{amssymb}
\usepackage{amsthm}
\usepackage{microtype}
\usepackage[retainorgcmds]{IEEEtrantools}
\usepackage{dsfont}

\def\kcore{$k$-core}

\def\hkcore{heterogeneous $k$-core}

\def\er{Erd\H{o}s-R\'enyi}

\def\ie{\textit{i.~e.}}
\def\eg{e.~g.}

\def\q{\vec{q}}

\begin{document}

\title{Weak percolation on multiplex networks}
\author{Gareth J.~Baxter}
\affiliation{Department of Physics \& I3N, University of Aveiro,
  Portugal}
\author{Sergey N. Dorogovtsev}
\affiliation{Department of Physics \& I3N, University of Aveiro,
  Portugal}
\affiliation{A. F. Ioffe Physico-Technical Institute, 194021
  St. Petersburg, Russia}
\author{Jos\'e F. F. Mendes}
\affiliation{Department of Physics \& I3N, University of Aveiro,
  Portugal}
\author{Davide Cellai}
\affiliation{MACSI, Department of Mathematics and Statistics, University of Limerick, Ireland}

\begin{abstract}
Bootstrap percolation is a simple but non-trivial model. It has applications in many areas of science and has been explored on random networks for several decades.
In single layer (simplex) networks, it has been recently observed that bootstrap percolation, which is defined as an incremental process, can be seen as the opposite of pruning percolation, where nodes are removed according to a connectivity rule.
Here we propose both a new model of bootstrap and of pruning percolation for multiplex networks.
We collectively refer to these two models with the concept of ``weak'' percolation, to distinguish them from the somewhat classical concept of ordinary (``strong'') percolation.
While the two models coincide in simplex networks, we show that they decouple when considering multiplexes, giving rise to a wealth of critical phenomena.
Our bootstrap model constitutes the simplest example of a contagion process on a multiplex network and has potential applications in critical infrastructure recovery and information security.
Moreover, we show that our pruning percolation model may provide a way to diagnose missing layers in a multiplex network.
Finally, our analytical approach allows us to calculate critical behavior and characterize critical clusters.
\end{abstract}

\date{\today}
\maketitle

\section{Introduction}

While network representations of complex systems have proven to be
tremendously useful, it is often the case that a single (simplex)
network cannot capture the complex interactions between systems or
sub-systems. Examples include  financial
\cite{caccioli2012,Huang2013}, ecological \cite{Pocock2012},
infrastructure \cite{Rinaldi2001} and information systems
\cite{leicht2009}.

Several different types of multilayer networks have  been introduced
in the past few years (for a review, see {\eg} \cite{kivela2013}).
In interdependent networks, corresponding nodes on different layers
may be linked by special dependency links, meaning the survival of a
node in one layer depends on the survival of its partner in another
layer \cite{Buldyrev2010}.
The nature of the inter-layer edges may
have different properties (dependence, control, etc.) and nodes may
not have a corresponding node on every other layer.
In multiplex networks, instead, the same nodes exist in every layer
and only the types of edges characterize the different layers
\cite{dedomenico2013,kivela2013,bianconi2013}.
For one-to-one interdependency, multiplex and interdependent networks are
equivalent with regard to percolation \cite{Son2011}.

These interdependencies between layers can have a profound effect on the
behavior of the entire system, behavior which could not be predicted
by studying each network in isolation. 
In particular, in interdependent or multiplex networks, damage to one
layer can spread to other layers, leading to a dramatic collapse of
the whole system \cite{Buldyrev2010,Buldyrev2011}.
Typically a
discontinuous hybrid phase transition is
observed \cite{baxter2012}, in contrast to the continuous transition
seen in classical percolation on a simplex network. 

Multiplexity ought to have effects on other network processes too.
In this paper we introduce an activation model on multiplex networks,
inspired by the bootstrap percolation process on a simplex
network. This represents activation of vertices on a network, such as
in social mobilisation, or the repair of infrastructure networks after
a disaster \cite{Duenas2007,Lee2007}. We also define its counterpart
pruning process.

%

The ordinary percolation process can be viewed equally as a damage or
as an activation process, and the result is the same. However, as we
will see, in the case of multiplex (and by extension, interdependent
networks), activation of the network yields a very different phase
diagram than a pruning/damage process. We describe a pair of
processes, which we call Weak Bootstrap Percolation (WBP) and Weak
Pruning Percolation (WPP), which represent activation/repair and
deactivation/damage processes respectively.
We refer to such two models as \emph{weak} to distinguish them from the simple (strong) extensions of classical percolation, where nodes belong to the same cluster if they are connected by homogeneous paths on each layer \cite{baxter2012}.
Incidentally, for strong  percolation, activation and pruning result in the same giant
  percolating cluster.
We also introduce the concept of invulnerable vertices, which are a
special category of vertices that are considered to be always
active. A small number of these vertices are necessary to stimulate
the activation of other vertices in the bootstrapping process. The
proportion of seed or invulnerable nodes also affects the nature of
the critical transitions observed. 

 Numerous generalizations of the original interdependent and multiplex
 models have appeared, examining the effects of reduced coupling
 strength \cite{Parshani2010}, link overlap
 \cite{hu2013,cellai2013c}, partial interdependence
 \cite{Gao2011,son2012}, and degree correlations \cite{min2013,valdez2013}
 among many others. Another theoretical approach involve spectral
 analysis of interconnected networks \cite{Radicchi2013}.
 Nevertheless, the focus has remained largely on the effects of damage
 on such networks. 
The present work is inspired by the bootstrap
 percolation process \cite{chalupa1979, BDG1} on a simplex network, which
 has recently been shown \cite{baxter2011} to be related to the {\kcore}
 process. We present a pair of multiplex models which exhibit complex
 critical behavior similar to that found in these simplex network
 models.
 
In Sec.~\ref{sec:models}, we introduce the models and describe the formalism for a multiplex network with an arbitrary number of layers. In Sec.~\ref{sec:wpp}, we solve the WPP model on {\er} network topologies on two and three layers, while in Sec.~\ref{sec:wbp} we do the same for the WBP model on two {\er} layers.
In Sec.~\ref{sec:clusters}, we characterize the critical clusters on both models and show that their size diverge at the transition and in Sec.~\ref{sec:conclusions} we present our conclusions.

\section{The models}
\label{sec:models}

\subsection{Formalism}

Before introducing the different models, we review the general
notation we are going to use in this paper. We consider a multiplex
network with multidegree distribution $P_{\q}$, where $\q=(q_1, \dots,
q_M)$.
There is growing tendency in the multiplex literature to indicate
layers with a  superscript and vertices with subscripts, for example
$q^m_i$ is the degree of type $m$ edges of vertex $i$ \cite{bianconi2013}, or using Greek letters for the vertices and Latin letters for the layer \cite{dedomenico2013}. 
In this paper, however, as we will rarely use
the vertex indices, we will only use subscript (Latin) indices to avoid
confusion with exponents.
Once the multidegree distribution is defined, the multiplexes we
consider are totally random, therefore they have the tree-like
property and vanishing edge overlap in the limit of large network size
\cite{bollobas2001}.

\begin{figure}[htb]
\begin{center}
	\includegraphics[width=0.5\columnwidth]{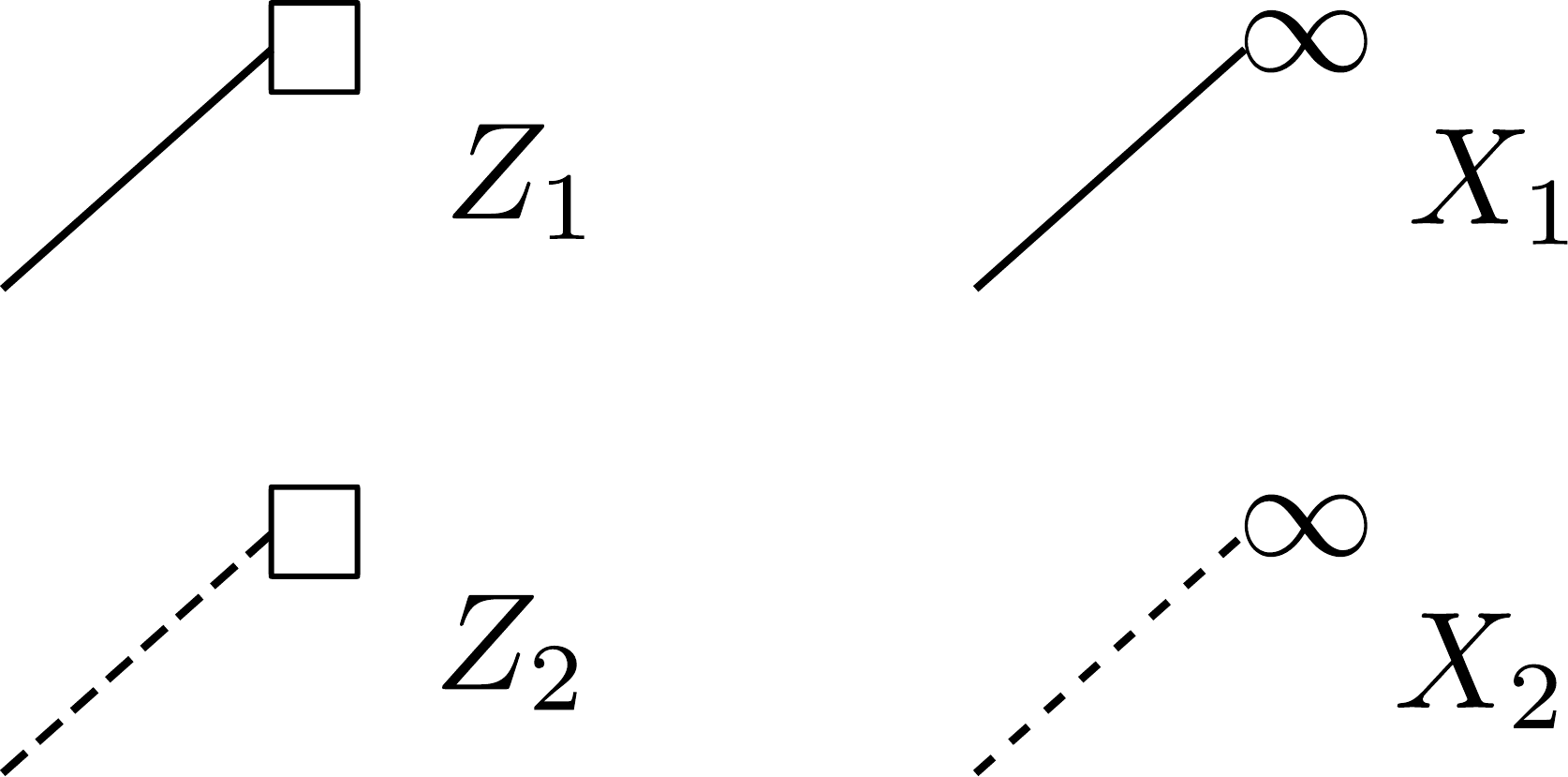}
	\caption{Diagrammatic representation of the probabilities
          $Z_n$ and $X_n$ in the case of two types of edges, $1$ and $2$.}
	\label{fig:diag-Z-X}
\end{center}
\end{figure}

In the following, we will consider percolation models that can be
defined by a pruning or a bootstrap process where vertices are
recursively removed or activated if their neighborhood fulfills a certain
property $\mathcal{P}$.
In addition, we allow a fraction $f$ of randomly selected vertices to be
considered part of the clusters at the end of the process even if they do
not satisfy the property $\mathcal{P}$ (invulnerable or seed vertices).
Then, we define the probability $S$ (often called \emph{strength} in
the classical percolation literature) that a randomly chosen vertex is
in the resulting largest cluster at the end of the process.
We will see that $S$ can be written in terms of two sets of
probabilities $\{Z_n\}$ and $\{X_n\}$.
$Z_n$ is simply the probability that, following an edge of type $n$,
the node encountered is in one of the resulting clusters at the end of
the process.
$X_n$ is the probability that, following an edge of type $n$, the node
encountered is in the \emph{infinite} resulting cluster at the end of
the process.
It is often instructive to represent equations for these probabilities
in graphical form. We therefore represent these variables by the
symbols shown in Fig.~\ref{fig:diag-Z-X}.
Finally, we also allow the multiplex network to be randomly damaged with probability $(1-p)$.
Here damage means that a fraction $(1-p)$ of the vertices, irrespective of their being among the $f$ invulnerable/seed nodes or not, are initially removed from the multiplex network together with their edges, of any type.
Our aim is to study the behavior of $S$ as a function of $p$ and describe the different critical phenomena that may arise in the different models.

\subsection{Strong Percolation (SP)}

An extension of percolation to multiplex networks has been
already studied \cite{Buldyrev2010,baxter2012}. In this paper, we refer to it as
\emph{strong percolation} (SP) to distinguish it from the new (weak)
percolation models we are going to define. For orientation, we briefly
recapitulate this percolation model, and introduce the concept of
invulnerable nodes.
A straightforward way to represent classical percolation on a
multiplex network is as a pruning process.
First, we assume that a fraction ($1-p$) of nodes is randomly removed.
Then, we recursively remove a
node if at least one of its degrees $q_m$ is zero.
As introduced in the previous section, we consider two kinds of
vertices: some are invulnerable with probability $f$, and some are
vulnerable with probability $(1-f)$.
Only vulnerable vertices can be pruned according to the rule above,
the invulnerable ones can always help in building a percolating
cluster as soon as they are connected to it by any edge type.
We define a mutually connected component as a cluster where each pair
of vertices is joined by a full path of edges of each type.
If the largest mutual connected component is infinite, we say that
there is strong percolation (SP).

Let $Z_n$ be the probability that, following an edge of type $n$, the
node encountered is part of a cluster where vertices can be removed no
further. The following equation holds:
\begin{equation}\label{eq:Z_SP}
	Z_{n} = p f + p(1-f) \sum_{\q}\frac{q_n P_{\q}}{\langle q_n\rangle} \prod_{
		\begin{subarray}{c}
        			m=1\\
			m\neq n
      		\end{subarray}}^M \left[ 1 - (1-Z_m)^{q_m}\right].
\end{equation}
The first term on the right hand side is the probability that the
encountered node is invulnerable ($f$) and undamaged ($p$).
The second term calculates the probability that, if the encountered
node is vulnerable, it is connected to an unpruned cluster by at least
one edge of each type.

At the end of the pruning process, the remaining clusters can be
finite or infinite, thus we define $X_n$ as the probability that,
following an edge of type $n$, the node encountered is attached to an
infinite mutually connected component by one of the outgoing edges and
has at least one adjacent edge of each type (for edges of type $n$,
the incoming edge is sufficient).
$X_n$ satisfies the following equation:
\begin{equation}
	X_{n} = p \sum_{\q}\frac{q_n P_{\q}}{\langle q_n\rangle}
        \left[ 1\! -\! (1\!-\!X_n)^{q_n-1}\right]\!
        \prod_{
	  \begin{subarray}{c}
            m=1\\
	    m\neq n
      	\end{subarray}}^M \!\left[ 1 - (1-X_m)^{q_m}\right].
\end{equation}
The right hand side calculates the probability that the encountered
node is connected to an unpruned infinite cluster by at least one edge
of each type.
Finally, the probability $S$ that a randomly chosen node is in the infinite percolating cluster is:
\begin{equation}
	S = p \sum_{\q} P_{\q} \prod_{m=1}^M \left[ 1 - (1-X_m)^{q_m}\right].
\end{equation}
It is easy to notice that when all nodes are vulnerable ($f=0$), there are not finite surviving clusters ($X_n = Z_n$), similarly to what occurs in {\kcore} percolation \cite{dorogovtsev2006,baxter2011}.

Alternatively, one could consider a bootstrapping process. 
Invulnerable nodes in the pruning process correspond to seed nodes in the bootstrap scheme.
Any node which has at least one connection of
each type to an active (percolating) cluster is added to the cluster. This
process is repeated until no more nodes can be added.
A strong percolating cluster requires a full path for each edge type
connecting each pair in the cluster \cite{baxter2012}. Since this is
a nonlocal property, in this model there is no distinction if the
network is pruned or bootstrapped, as the difference between the two
definitions is local, thus it does not affect the requirement of an
infinite cluster for each edge type.

\subsection{Weak Pruning Percolation (WPP)}

We now define a multiplex percolation process that is entirely local 
and, as we will see, gives rise to distinct processes when we
consider either pruning or bootstrapping.

Let us consider a multiplex network with multidegree distribution $P_{\q}$.
Vertices are randomly assigned, with probability $f$,
the property of being invulnerable, the remaining ones being
vulnerable instead.
We consider a pruning process where only vulnerable vertices can be pruned.
More specifically, we define Weak Pruning Percolation (WPP) as the
process in which every vulnerable node 
in a multiplex network is
recursively pruned if at least one of its degrees 
$q_m$
is zero.

Let $Z_n$ be the probability that, following an edge of type $n$, the
node encountered is part of a cluster where vertices can be removed no
further.
At the end of the pruning process, the remaining clusters can be
finite or infinite, thus we define $X_n$ as the probability that,
following an edge of type $n$, the node encountered is attached to an
infinite cluster by one of the outgoing edges and has at least one
adjacent edge of each type (for edges of type $n$, the incoming edge
is sufficient).
Differently from SP, this definition does not require that the encountered node is
connected to an infinite cluster by every edge type, but only that
there exist at least one outgoing edge to an infinite cluster at each
step.
That is why we call this model \emph{weak} pruning percolation.

In a multiplex with $M$ types of edges and a degree distribution
$P_{\vec{q}}$, we can write equations for the variables $Z_n$ and
$X_n$ for the generic edge type $n$.
The equation for $Z_n$ is:
\begin{equation}
	Z_{n} = p f + p (1-f) \sum_{\vec{q}} \frac{q_n
          P_{\vec{q}}}{\langle q_n\rangle}\prod_{
	  \begin{subarray}{c}
            m=1\\
	    m\neq n
      	  \end{subarray}
	}^{M}\left[1 - (1-Z_m)^{q_m} \right].
	\label{eq:Z-wpp-general}
\end{equation}
As $Z_n$ represents the probability that following an edge of type $n$
an unpruned cluster is reached, this equation consists of two terms.
The first term ($pf$) accounts for the probability that the encountered node is an undamaged invulnerable node.
The second term (proportional to $p(1-f)$) calculates the corresponding probability of not being pruned for a vulnerable node.
It is calculated as the product of probabilities that at least one neighbor by each edge type is in an unpruned cluster.
We do not need to consider the same edge type ($n$) we have picked, as
the very existence of that edge implies an unpruned neighbor at the
other end of the edge we are considering.
Note that equation (\ref{eq:Z-wpp-general}) is identical to the
equivalent equation for strong percolation, (\ref{eq:Z_SP}). We repeat
it here to emphasize that these are distinct models.

\begin{figure*}[htb]
	\includegraphics[width=0.95\textwidth,angle=0]{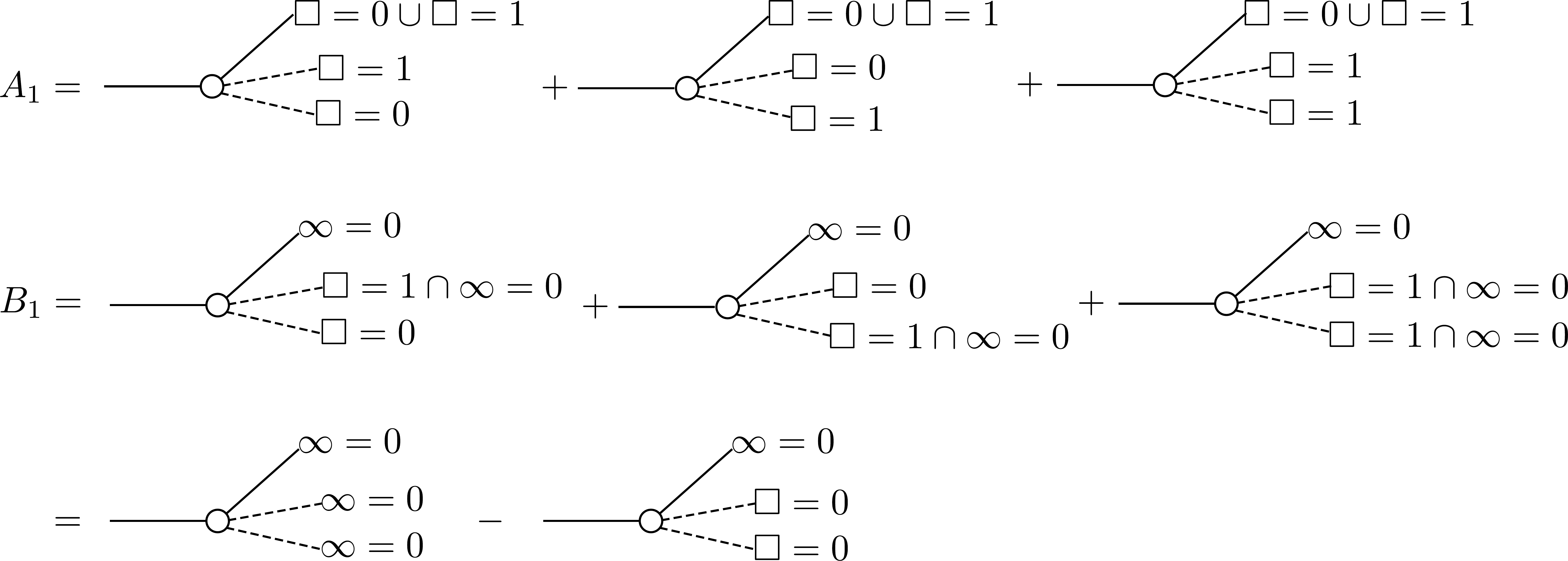}
	\caption{Diagrams describing the second term on the right hand side of equation (\ref{eq:X-wpp-general}) in the case of a node with multidegree  $(q_1, q_2) = (2, 2)$:
	$A_1 = [1-(1-Z_2)^{q_2}] $ and $B_1 = (1-X_1)^{q_1-1} [(1-X_2)^{q_2}-(1-Z_2)^{q_2}] $.
	$A_1$ calculates the probability that, after having followed a type-1 edge, we get to a node from which at least one edge of type 2 leads to an unpruned cluster (finite or infinite).
	We can display this as a sum of three terms representing all the relevant possibilities in the case of $q_2=2$.
	$B_1$ is composed of two factors.
	The first factor represents the probability that no edge of type 1 lead to an infinite unpruned cluster ($\infty = 0$).
	The second factor calculates the probability that all the edges of type 2 either lead to finite unpruned clusters ($\square = 1 \cap \infty = 0$) or no clusters at all ($\square = 0$).
	This is the meaning of the first line of $B_1$.
	The second line shows that $B_1$ can be written in a more compact way as a difference between the probability that none of the outgoing edges of type 2 lead to an infinite cluster and the probability that all the outgoing edges of type 2 do not lead to an unpruned cluster (even finite).
	This second line explains pictorially the way we write $B_1$ in equation (\ref{eq:X-wpp-general}).
	}
	\label{fig:wpp-example-diag-X}
\end{figure*}

The equation for $X_n$, instead, is
\begin{multline}
    X_{n} = p f  \sum_{\vec{q}} \frac{q_n P_{\vec{q}}}{\langle q_n\rangle} \big[ 1 - (1-X_n)^{q_n-1} \prod_{
        \begin{subarray}{c}
                    m=1\\
            m\neq n
              \end{subarray}}^{M}
    (1-X_m)^{q_m} \big] \\
 + p(1-f) \sum_{\vec{q}}\frac{q_n P_{\vec{q}}}{\langle q_n\rangle}
\Bigg\{ \prod_{
        \begin{subarray}{c}
                    m=1\\
            m\neq n
              \end{subarray}}^{M}
    [1-(1-Z_m)^{q_m}] \\
 -  (1\!-\!X_n)^{q_n\!-1} \prod_{
        \begin{subarray}{c}
                    m=1\\
            m\neq n
              \end{subarray}}^{M}
    [(1-X_m)^{q_m}-(1-Z_m)^{q_m}] \Bigg\}.
    \label{eq:X-wpp-general}
\end{multline}
The first sum on the right hand side calculates the probability that the type-$n$ edge we are following encounters an invulnerable node which is attached to an infinite cluster which cannot be pruned any further.
The second sum calculates the same probability but in the case when the encountered node is not invulnerable.
This term is written as a difference between the probability of having at least one finite unpruned cluster by each edge type and another term which removes the possibility that for any edge type all the unpruned clusters are finite.
This second product must be multiplied by $(1-X_n)^{q_n-1} $ to exclude the possibility that an infinite percolating cluster is accessible by a type $n$ edge.
The rationale of this calculation is illustrated in Fig.~\ref{fig:wpp-example-diag-X}, which displays the two terms of this difference in a simple example.
Finally, having given equations for $Z_n$ and $X_n$, we can use them
to find $S$, the probability that a randomly chosen
node is in the giant percolating cluster defined in this model. This
is the strength of the giant percolating cluster.
It is given by the following formula:
\begin{eqnarray}
	S =& p f  \sum_{\vec{q}} P_{\vec{q}} \left[ 1 -  \prod_{m=1}^{M}
	(1-X_m)^{q_m} \right]  \nonumber\\
	  & + p (1-f) \sum_{\vec{q}} P_{\vec{q}} \bigg\{  \prod_{m=1}^{M} \left[ 
	1 - (1-Z_m)^{q_m} \right]\nonumber\\
        & - \prod_{m=1}^{M}  \left[ 
	  (1-X_m)^{q_m} - (1-Z_m)^{q_m} \right]\bigg\}.
	\label{eq:S-wpp-general}\label{eq:S-wbp-general}
\end{eqnarray}
The first term calculates the probability that  an invulnerable node is part of the giant weak pruning cluster by subtracting from one the case where each neighbor of the considered node is not part of the giant weak pruning cluster.
The second term calculates the same probability in the case of vulnerable nodes.
In particular, the first product evaluates the probability that at least one neighbor of each type is in a weak pruning cluster.
From that we have to subtract the second product, which corresponds to the probability that at least one neighbor of a type belongs to a finite weak pruning cluster.

\subsection{Weak Bootstrap Percolation (WBP)}

In simplex networks, a bootstrap process is generally defined by a
simple contagion mechanism where a node becomes active as soon as at
least $k$ of its adjacent nodes is active \cite{chalupa1979}. When
$k=1$ this corresponds to ordinary percolation. Bootstrap percolation
occurs when a giant fraction of the system becomes active at the end
of the process.
We now propose an extension of this model to multiplex
networks. A node becomes active when it has $M$ active neighbors, one
in each layer of the multiplex.
Let us consider that vertices are initially active (seed) with
probability $f$, inactive with probability $1-f$.
We define \emph{weak bootstrap} as the process where every
inactive node in a multiplex network is activated if at least one of
the neighbors connected by each edge type is active.
We say that Weak Bootstrap Percolation (WBP) occurs if a finite fraction of nodes is activated at the end of the process.
In addition, we allow a fraction ($1-p$) of the multiplex to be randomly damaged and study how the behavior of the percolating cluster depends on the parameters $f$ and $p$.
As in the simplex bootstrap scheme, this process is monotonic, {\ie} active vertices cannot become inactive.

\begin{figure}[htb]
	\includegraphics[width=0.99\columnwidth]{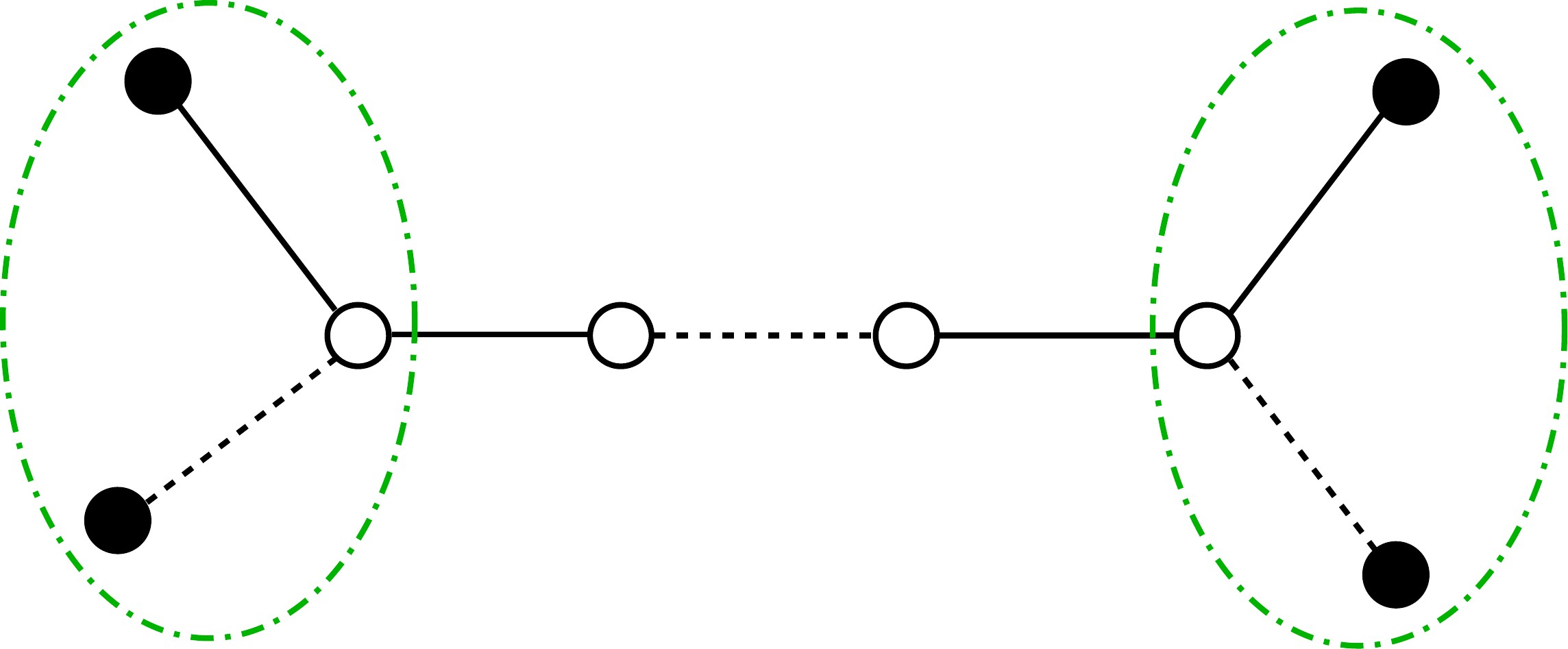}
	\caption{Example of clusters in a multiplex with two types of edges.
	Black nodes are invulnerable/seed vertices, white nodes are vulnerabl vertices.
	In WPP, all the nodes are unprunable (remain active), because each white node
        is connected to another node by each edge type.
	In WBP, only the nodes inside the green dot-dashed lines become active, while the remaining two nodes have only one active neighbor, by one edge type only, so they cannot become active.}
	\label{fig:difference-wpp-wbp}
\end{figure}

At the end of the activation process, the active clusters are in
general not the same as those that would be found through the pruning
process. This is because in WPP nodes are considered active until
pruned. This means that, for example, a pair of nodes connected by an edge
of one type, provide the required support of that type for one
another, even if neither has another edge of that type. In WBP, on the
other hand, such an isolated dimer can never become activated (Fig.~\ref{fig:difference-wpp-wbp}). The
same holds for many larger configurations as well.

At the end of the activation process, let $Z_n$ be the probability
that, following an edge of type $n$, the node encountered is part of a
cluster of active vertices.
\begin{multline}
	Z_{n} = p f + p (1-f) \sum_{\vec{q}} \frac{q_n
          P_{\vec{q}}}{\langle q_n\rangle} \left[ 1 -
          (1-Z_n)^{q_n-1}\right] \\ 
        \times \prod_{	\begin{subarray}{c}
            m=1\\
	    m\neq n
      	\end{subarray}}^{M}
	\left[1 - (1-Z_m)^{q_m} \right] .
	\label{eq:Z-wbp-general}
\end{multline}
The first term in this equation ($pf$) accounts for the probability that the encountered node is an undamaged seed node.
The second term (proportional to $p(1-f)$) calculates the corresponding probability of being active for a non-seed node.
It is calculated as the product of probabilities that at least one neighbor by each edge type is in an active cluster.
In the case where we are considering the same edge type we have picked, a further active node must exists by an edge different from the one we came from.

Such clusters of active vertices can be finite or infinite, thus we
define $X_n$ as the probability that, following an edge of type $n$,
the node encountered is attached to an infinite cluster by one of the
outgoing edges and has at least one outgoing edge of each type (even
for edges of type $n$, the incoming edge is \emph{not} sufficient,
unlike the WPP model).
This definition does not require that the encountered node is
connected to an infinite cluster by each edge type, but only that
there exist at least one outgoing edge to an infinite cluster at each
step. That is the reason why we call this model \emph{weak} bootstrap
percolation.

\begin{figure*}[htb]
	\includegraphics[width=0.95\textwidth,angle=0]{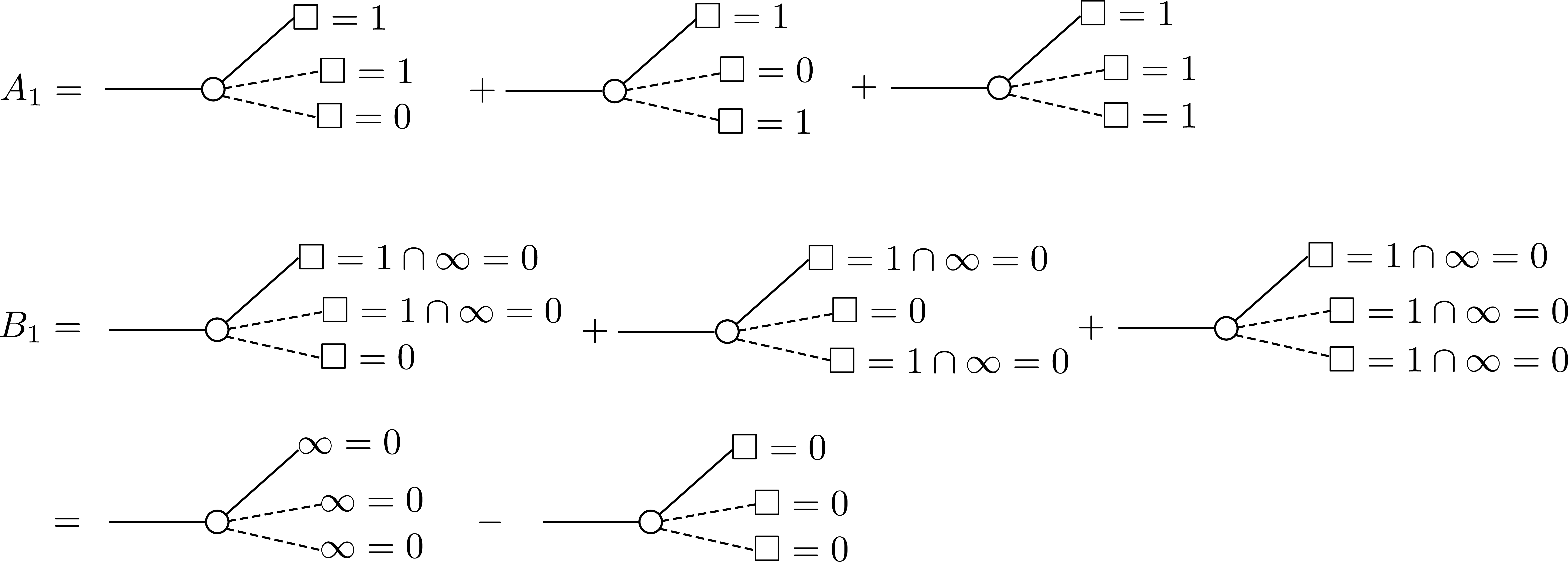}
	\caption{Diagrams describing the second term on the right hand side of equation (\ref{eq:X-wbp-general}) in the case of a node with multidegree  $(q_1, q_2) = (2, 2)$:
	$A_1 = [1-(1-Z_1)^{q_1-1}][1-(1-Z_2)^{q_2}] $ and $B_1 =  [(1-X_1)^{q_1-1}-(1-Z_1)^{q_1-1}] [(1-X_2)^{q_2}-(1-Z_2)^{q_2}] $.
	$A_1$ calculates the probability that, following an edge of type 1, we get to a node  from which at least one edge of type 2 leads to an active cluster (finite or infinite).
	We can display this as a sum of three terms representing all the relevant possibilities in the case of $q_2=2$.
	$B_1$ is composed of two factors, one for each type of edges.
	In the first line, we show that each factor calculates the probability that all the edges of type 1 or 2 either lead to finite active clusters ($\square = 1 \cap \infty = 0$) or no clusters at all ($\square = 0$).
	This can be written in a more compact way (second line) as a difference between the probability that none of the outgoing edges of type 2 lead to an infinite cluster and the probability that all the outgoing edges of type 2 do not lead to an active cluster (even finite).
	This second line explains pictorially the way we write $B_1$ in equation (\ref{eq:X-wbp-general}).
	}
	\label{fig:wbp-example-diag-X}
\end{figure*}

An argument similar to the one regarding Eq.~(\ref{eq:X-wpp-general})
yields the following equation for $X_n$:
\begin{multline}
	X_{n} = p f  \sum_{\vec{q}} \frac{q_n P_{\vec{q}}}{\langle q_n\rangle} \bigg[ 1 - (1-X_n)^{q_n-1} \prod_{
		\begin{subarray}{c}
        			m=1\\
			m\neq n
      		\end{subarray}}^{M}
	(1-X_m)^{q_m} \bigg] \\
+ p(1-f) \sum_{\vec{q}}\frac{q_n P_{\vec{q}}}{\langle q_n\rangle}
\bigg\{ [ 1 - (1-Z_n)^{q_n-1}] \prod_{
		\begin{subarray}{c}
        			m=1\\ 
			m\neq n
      		\end{subarray}}^{M}
	[1-(1-Z_m)^{q_m}] \\
 - [(1-X_n)^{q_n-1} - (1-Z_n)^{q_n-1}] \\ \times
\prod_{
		\begin{subarray}{c}
        			m=1\\ 
			m\neq n
      		\end{subarray}}^{M}
	[(1-X_m)^{q_m}-(1-Z_m)^{q_m}] \bigg\}.
	\label{eq:X-wbp-general}
\end{multline}
%
The meaning of this equation is schematically explained in Fig.~\ref{fig:wbp-example-diag-X}.
Similarly to Eq.~(\ref{eq:X-wpp-general}), the first term calculates the probability that following a randomly chosen $n$-type edge we get to a seed node where at least one of the outgoing neighbors is part of a giant activated cluster.
The second term contains the difference of two products.
The first product represents the probability of having at least one activated cluster by each edge type.
Then, we have to subtract to this quantity the second product, which calculates the probability that, for any edge type, all the activated clusters are finite.
The second sum in equation (\ref{eq:X-wbp-general}) is schematically exemplified in Fig.~\ref{fig:wbp-example-diag-X}.

While $Z_n$ and $X_n$ are different from their WPP counterparts, the
equation for $S$ is the same as Eq. (\ref{eq:S-wpp-general}).
\section{WPP on {\er} networks}
\label{sec:wpp}

To demonstrate the qualitative behavior of these models, we apply the
formalism described above for WPP to uncorrelated {\er} networks. In
the following Section we will repeat these calculations for WBP.

\subsection{Two identical {\er} networks}

\begin{figure}[htb]
\begin{center}
	\includegraphics[width=0.9\columnwidth]{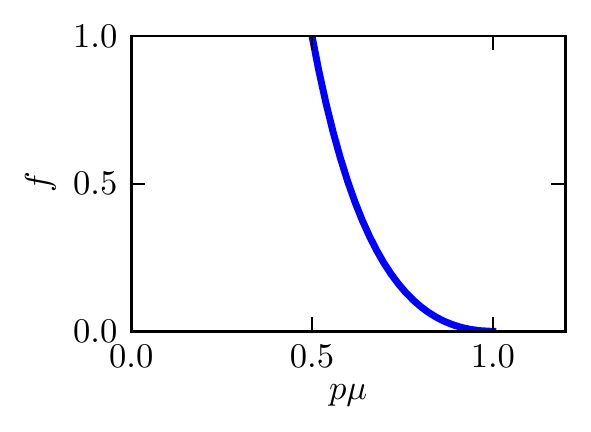}
	\caption{Phase diagram of the WPP model for two uncorrelated {\er} networks with identical mean degree $\mu$.}
	\label{fig:wpp-ph-diag}
\end{center}
\end{figure}

We first apply the formalism to uncorrelated {\er} networks with
identical mean degree $\mu$.
The symmetry of the identical layers implies that
$Z_1=Z_2\equiv Z$ and $X_1=X_2\equiv X$.

\begin{equation}
	Z = p \left[ 1 - (1-f) e^{-\mu Z}\right],
\end{equation}
\begin{equation}
	X = p \left[ 1 - e^{-2\mu X}  -(1-f)e^{-\mu Z} (1-e^{-\mu X})  \right]\,.  
\end{equation}
These equations can be rescaled with the variables $x = X/p$, $z = Z/p$, $\nu=p\mu $ and rewritten as
\begin{equation}
	\Phi_{f,\nu} (z) =  \frac{1 - (1-f) e^{-\nu z}}{z}  = 1\,,
	\label{eq:wpp-symm-cont-z}
\end{equation}
\begin{equation}
	\Psi_{\nu} (x)  = \frac{ 1 - e^{-2\nu x} -x }{ 1 - e^{-\nu x}
        } = (1-f)  e^{-\nu z} \,.
	\label{eq:wpp-symm-cont-x}
\end{equation}
As $\Phi_{\nu}(z)$ and $\Psi_{\nu}(x)$ are always monotonically decreasing in $z$ and $x$, respectively, the phase diagram is characterized by a line of continuous phase transitions when 
\begin{equation}
	\Psi_{\nu} (0) = 2 - \frac{1}{\nu_c} = 1 - z_c,
	\label{eq:wpp-symm-cont-x2}
\end{equation}
which can only occur when $\nu_c\le 1$.
Therefore, the line of continuous transitions is given by substituting (\ref{eq:wpp-symm-cont-x2}) into (\ref{eq:wpp-symm-cont-z}):
\begin{equation}
	f_c = 1- e^{1-\nu_c} \left( 2 - \frac{1}{\nu_c}\right),
\end{equation}
with $\nu_c\le 1$.
The phase diagram of the model is shown in Fig.~\ref{fig:wpp-ph-diag}.
At $f=0$ (every node is vulnerable), we find the  condition of classical percolation $p\mu = 1$.
As there are no correlations between layers, the probability that a randomly chosen node is in the WPP giant cluster corresponds to the probability that it is in the giant component in each layer.
In the case of identical mean degrees, that occurs at the same point $p\mu = 1$.
At $f=1$, instead, every node is invulnerable, so the network behaves as a simplex network where the edge types are indistinguishable.
The model becomes then equivalent to a classical percolation scenario of an {\er} network with effective mean degree $2\mu$.

\subsection{Two non-identical {\er} networks}

When two uncorrelated {\er} networks have different mean degrees
$\mu_1$ and $\mu_2$, equations as (\ref{eq:Z-wpp-general}) become
\begin{eqnarray}
	Z_1 &= p \left[ 1 - (1-f) e^{-\mu_2 Z_2}\right]\\
        Z_2 &= p \left[ 1 - (1-f) e^{-\mu_1 Z_1}\right]
\end{eqnarray}
while (\ref{eq:X-wpp-general}) becomes
\begin{eqnarray}
	X_1 &= p \left[ 1\! - e^{-\mu_1 X_1-\mu_2X_2}  -(1\!-\!f)e^{-\mu_2 Z_2} (1\!-\!e^{-\mu_1 X_1})
          \right]\nonumber\\
          &\\
        X_2 &= p \left[ 1\! - e^{-\mu_1 X_1-\mu_2X_2}  -(1\!-\!f)e^{-\mu_1 Z_1} (1\!-\!e^{-\mu_2 X_2})
          \right]. \nonumber\\
          &
\end{eqnarray}
These equations can be rescaled with the variables $x_1 = X_1/p$, $x_2 = X_2/p$, $z_1 = Z_1/p$, $z_2 = Z_2/p$, $\nu_1=p\mu_1 $, $\nu_2=p\mu_2 $ and rewritten as

\begin{equation}
z_1 = 1 - (1-f)e^{-\nu_2z_2}\,, \qquad z_2 = 1 - (1-f)e^{-\nu_1z_1} 
\end{equation}
which can be reduced to a single equation by substituting one into the
other. For $x_1$ and $x_2$ we have
\begin{eqnarray}
x_1 = z_1 - e^{-\nu_1x_1}(e^{-\nu_2x_2}+z_1-1)\label{eq:WPP_ER_z1}\,,\\
x_2 = z_2 - e^{-\nu_2x_2}(e^{-\nu_1x_1}+z_2-1)\label{eq:WPP_ER_z2}\,.
\end{eqnarray}

Assuming that there is a continuous transition in $X_1$, and that the
transition point $f_c$ is the same for both networks, these become, in
the limit $x_1, x_2 \to 0$,
\begin{equation}
1 = \nu_2 \frac{x_2}{x_1} + \nu_1z_1\,, \qquad 1 = \nu_1
\frac{x_1}{x_2} + \nu_2z_2\,. 
\end{equation}
Eliminating $x_1/x_2$, we find that at the critical point
\begin{equation}\label{eq:WPP_ER_nu_cp}
\nu_1\nu_2 = (1-\nu_1z_1)(1-\nu_2z_2)\,.
\end{equation}
Simultaneous solution of (\ref{eq:WPP_ER_z1}),(\ref{eq:WPP_ER_z2}) and
(\ref{eq:WPP_ER_nu_cp}) allows us to find the line of the transition.
In the limit $f=0$, the solution is $z_1=z_2=0$ and $\nu_1\nu_2 =
1$.
As in the identical mean degree case, in this limit the probability of a node being in the giant WPP component is given by the product of the classical percolation probability in each layer.
Therefore, even if one layer does not percolate on its own ($\nu_1<1$), the other one may provide more edges to support the weak pruning percolating cluster ($\nu_2 = 1/\nu_1 > 1$).
In the other limit $f=1$, we find $z_1 = z_2 = 1$ and $\nu_1 + \nu_2 =
1$. Note that this corresponds to the classical percolation threshold,
if the multiplex is treated as a single network with mean degree
$\mu_1+\mu_2$ (no distinction between kinds of edges).
Intermediate values can be found by numerical solution.
\begin{figure}[htb]
\begin{center}
	\includegraphics[width=0.95\columnwidth]{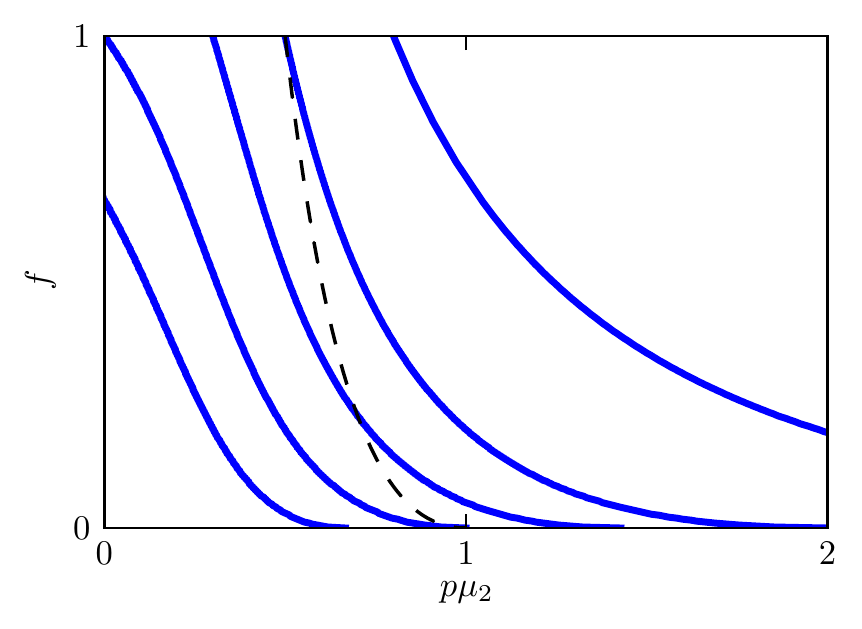}
	\caption{Phase diagram of the WPP model for two uncorrelated
          {\er} networks with mean degree $\mu_1$ and
          $\mu_2$. Horizontal axis is $\nu_2 = p\mu_2$. Each solid
          curve shows the location of the continuous transition for a
          different value of $\nu_1 = p\mu_1$, from right to left
          $\nu_1 = \{0.2,0.5,0.7,1,1.5\}$. The black dashed line
          represent points where $\mu_1 = \mu_2$.}
	\label{fig:wpp-ph-diag_ERasym}
\end{center}
\end{figure}

\subsection{Three identical {\er} networks}
\label{sec:WPP-3layer}

As there is no discontinuous transition in the 2-layer case, to
show that such a transition can indeed occur in the WPP model, we
now consider WPP on three {\er} networks with identical mean
degree $\mu$. From equations (\ref{eq:Z-wpp-general}) and
(\ref{eq:X-wpp-general}), and observing that from the symmetry of the
problem we have $Z_1=Z_2=Z_3$ and $X_1=X_2=X_3$, we have
\begin{equation}
	z = 1 - (1-f)  e^{-\nu z} \left( 2 - e^{-\nu z}\right),
	\label{eq:3layer-WPP-z}
\end{equation}
\begin{equation}
	x = 1 - e^{-3\nu x} - (1-f) e^{-\nu z} \left( 2 - e^{-\nu z}
        +e^{-\nu (z+x)} - 2e^{-2\nu x}  \right) \label{eq:3layer-WPP-x}
\end{equation}
\begin{equation}
	\frac{S}{p} = 1 - e^{-3\nu x} - 3 (1-f) e^{-\nu z} \left( 1 - e^{-\nu z} +e^{-\nu (z+x)} - e^{-2\nu x}  \right)
	\label{eq:3layer-WPP-S_ER}
\end{equation}
where $z=Z/p$, $x=X/p$ and $\nu = p\mu$.

As in the 2-layer case, the continuous transition can be found by imposing $x=0$, yielding
\begin{equation}
	\begin{cases}
		z = 1 - (1-f)  e^{-\nu z} \left( 2 - e^{-\nu z}\right) \\
		1 - 3\nu + 4\nu(1-f) e^{-\nu z} - \nu(1-f) e^{-2\nu z} = 0
	\end{cases}.
\end{equation}
Once again, for $f=1$ the system behaves as a simplex network with mean degree $3\mu$ and therefore the transition occurs at $p\mu = 1/3$.
At $f=0$, instead, the system collapses at any value of $p\mu$.
As can be seen analytically, the equations can be manipulated in the case of $f=0$ so that $\nu$ must satisfy the equation
\begin{multline}
	1 + \frac{1}{\nu}\ln\left( 2-\sqrt{1+\frac{1}{\nu}} \right)
        =\\
 \left( 2-\sqrt{1+\frac{1}{\nu}} \right) \left[ 2 - \left( 2-\sqrt{1+\frac{1}{\nu}} \right)^2 \right],
\end{multline}
which has no solution for $\nu > 0$.

\begin{figure}[htb]
\begin{center}
	\includegraphics[width=0.95\columnwidth,angle=0]{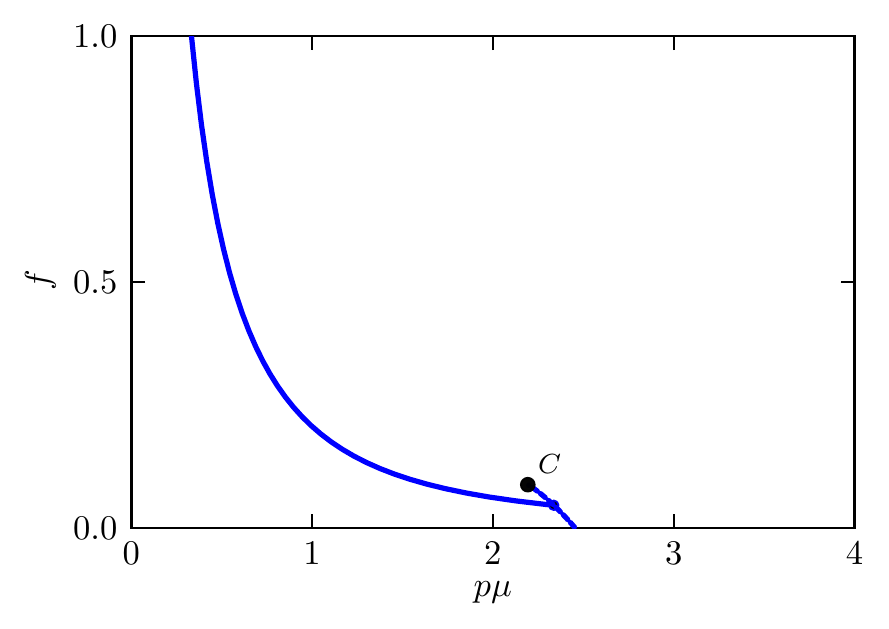}
	\caption{Phase diagram of the WPP model for three uncorrelated
          {\er} networks with identical mean degree $\mu$. The solid
          line gives the location of the continuous transition, the
          dashed line gives the location of the discontinuous
          transition. The point $C$ is the critical point.}
	\label{fig:wpp-3er-ph-diag}
\end{center}
\end{figure}

Unlike in the 2-layer case, however, here we also have a line of discontinuous transitions.
From equation (\ref{eq:3layer-WPP-z}), we can write $\Phi_{f,\nu} (z) = 1$, where
\begin{equation}
	\Phi_{f,\nu}(z)  = \frac{1 - (1-f)  e^{-\nu z} \left( 2 - e^{-\nu z}\right)}{z}.
	\label{eq:WPP-3layer-phi}
\end{equation}
This function is not monotonic in $z$, and therefore we have a critical point when $\Phi_{f,\nu}(z) = 1$, $\Phi^{\prime}_{f,\nu}(z) = \Phi^{\prime\prime}_{f,\nu}(z) = 0$ which yield
\begin{equation}
	f_{CP} = \frac{2\ln 2 -1}{2\ln 2 +3} = 0.088\dots
\end{equation}
\begin{equation}
	\nu_{CP} = \ln 2 +  \frac{3}{2} = 2.193\dots
\end{equation}
\begin{equation}
	z_{CP} = \frac{2\ln 2}{2\ln 2 +3} =  0.316\dots
\end{equation}
The expansion of $S$ (Eq. (\ref{eq:3layer-WPP-S_ER})) around the critical point at $f=f_c$ yields $S - S_c \sim (\nu - \nu_c)^{\beta}$ with $\beta = 1/3$, as in the critical points observed in simplex {\hkcore} percolation \cite{baxter2011,cellai2013b}.
The line of discontinuous transitions can be calculated by imposing the conditions
\begin{equation}
	\left\{%
		\begin{array}{lcr}
			\Phi_{f,\nu}(z) &=& 1\\
			\Phi^{\prime}_{f,\nu}(z) &=& 0\\
			\Phi^{\prime\prime}_{f,\nu}(z) &<& 0
		\end{array}
	\right.
\end{equation}
because we are looking for the point where the maximum of the function encounters the line at 1.
The line of continuous transitions intersects the line of discontinuous transitions at a triple point that can be found by imposing the conditions
\begin{equation}
	\left\{%
		\begin{array}{lcr}
			\Phi_{f,\nu}(z_d) &=& 1\\
			\Phi^{\prime}_{f,\nu}(z_d) &=& 0\\
			\Phi_{f,\nu}(z_c) &=& 1\\
			\Psi_{f,\nu}(0, z_c) &=& 1
		\end{array}
	\right.
\end{equation}
where $z_d$ is the value of $z$ at the discontinuous transition,
$z_c$ the value at the continuous transition, and $\Psi_{f,\nu}$ is defined, from the (\ref{eq:3layer-WPP-x}), as
\begin{multline}
	\Psi_{f,\nu}(x,z)= \frac{1}{x}\bigg[
 1 - e^{-3\nu x}\\
 - (1-f) e^{-\nu z} \left( 2 - e^{-\nu z} +e^{-\nu (z+x)} - 2e^{-2\nu
   x}  \right)\bigg].
\end{multline}
Those four equations
yield the position of the triple point: $f_{t} =  0.0462253\dots$, $\nu_{t}
= 2.33666\dots$. The phase diagram of this model is illustrated in
Fig.~\ref{fig:wpp-3er-ph-diag}.

Finally, we need to check if Eq.~(\ref{eq:3layer-WPP-x}) can have multiple solutions in $x$ at fixed $z$.
If that is the case, the model may be characterized by another discontinuous transition, not captured by Eq.~(\ref{eq:3layer-WPP-z}).
If a critical point exists, then it must satisfy the equations $\Psi_{f,\nu}(x,z)=1$, $\frac{\partial\Psi_{f,\nu}}{\partial x} = \frac{\partial^2\Psi_{f,\nu}}{\partial x^2} =0$, that yield
\begin{multline}
			1-x - e^{-3\nu x} \\
 - (1-f) e^{-\nu z} \left( 2 - e^{-\nu z} +e^{-\nu (z+x)} - 2e^{-2\nu x}
                        \right) = 0 \\
		\shoveleft{ 1-\nu (1-f) e^{-2\nu z}e^{-\nu x}}\\
  \shoveright{ + 4\nu (1-f) e^{-\nu z}e^{-2\nu x} - 3\nu e^{-3\nu x} = 0} \\
		\shoveleft (1-f) e^{-2\nu z} -8(1-f) e^{-\nu z}e^{-\nu
                  x} + 9 e^{-2\nu x} = 0
\end{multline}
From the third equation, we can work out a relationship between $x$ and $z$:
\begin{equation}
	x = z -\frac{1}{\nu}\ln\left[ \frac{1}{9} (1-f) \left(4\pm\sqrt{\frac{7-16f}{1-f}} \right) \right],
\end{equation}
but this is impossible because it implies that $x>z$ for any value of $f$.
Therefore, Eq.~(\ref{eq:3layer-WPP-x}) does not have solutions associated
with extra discontinuous phase transitions.

\section{WBP on {\er} networks}
\label{sec:wbp}

Now let us compare these results with the WBP model.

\subsection{Two identical {\er} layers}

First we consider two uncorrelated {\er} networks with identical mean
degree $\mu$. The symmetry of the identical layers implies that
$Z_1=Z_2$ and $X_1=X_2$, so that we can simply define two variable $Z
\equiv Z_1 = Z_2$ and $X \equiv X_1 = X_2$. Then
\begin{equation}
	Z = pf + p (1-f) \left( 1 - e^{-\mu Z}\right)^2
\end{equation}
and
\begin{equation}
	X = p \left( 1 - e^{-2\mu X}\right)  - p (1-f) 2 e^{-\mu Z}  \left(1-e^{-\mu X}\right).  
\end{equation}

These equations can be rescaled with the variables $x = X/p$, $z = Z/p$, $\nu=p\mu$ and rewritten as
\begin{equation}
	\Phi_{f,\nu} (z) =  \frac{1 - (1-f) e^{-\nu z}(2-e^{-\nu z})}{z}  = 1
\end{equation}
\begin{equation}
	\Psi_{\nu} (x)  = \frac{ 1 - e^{-2\nu x}  -x }{ 2 (1 - e^{-\nu x}) } = (1-f)  e^{-\nu z} 
\end{equation}

\begin{figure}[htb]
\begin{center}
	\includegraphics[width=0.95\columnwidth]{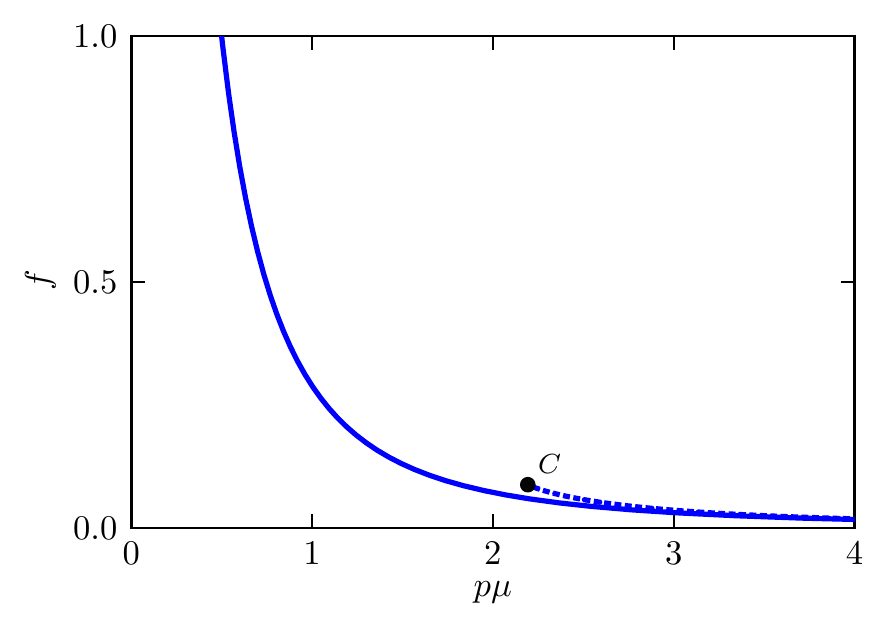}
	\caption{Phase diagram of the WBP model for two uncorrelated
          {\er} networks with identical mean degree $\mu$. The solid
          line gives the location of the continuous transition, the
          dashed line gives the location of the discontinuous
          transition. The point $C$ is the critical point.
	}
	\label{fig:wbp-ph-diag}
\end{center}
\end{figure}

Fig.~\ref{fig:wbp-ph-diag} displays the phase diagram of this model.
As $\Psi_{\nu} (x) $ is always a monotonic decreasing function, the maximum occurs at $\Psi_{\nu} (0) = 1-\frac{1}{2\nu}$.
Therefore, the following equation defines the line of continuous transitions between percolating ($x>0$) and a non-percolating phase ($x=0$):
\begin{equation}
	(1-f)  = e^{\nu z} \left(1 -\frac{1}{2\nu}\right).
	\label{eq:wbp-symm-cont}
\end{equation}

However, as the function $\Phi_{f,\nu} (z)$ is not always monotonic, the phase diagram also contains a line of discontinuous transitions.
This line is defined by the conditions
\begin{equation}
	\left\{%
		\begin{array}{lcr}
			\Phi_{f,\nu}(z) &=& 1\\
			\Phi^{\prime}_{f,\nu}(z) &=& 0\\
			\Phi^{\prime\prime}_{f,\nu}(z) &>& 0\,.
		\end{array}
	\right.
	\label{eq:wbp-symm-disc}
\end{equation}
The last condition captures the lower branch of the solutions (which is a minimum of $\Phi$), as the higher branch is unphysical in a bootstrap model \cite{baxter2011}.
In fact, the instable branch corresponds to the stable branch of the WPP model on three layers that we have calculated in Sec.~\ref{sec:WPP-3layer}.
The line of discontinuous transitions ends at a critical point defined by the conditions $\Phi_{f,\nu}(z) = 1$, $\Phi^{\prime}_{f,\nu}(z) = \Phi^{\prime\prime}_{f,\nu}(z) = 0$.
As $\Phi_{f,\nu}(z)$ is identical as in the WPP model on three layers (\ref{eq:WPP-3layer-phi}), the critical point is exactly the same and belongs to the same class of universality $\beta = 1/3$.

The discontinuous line tends to zero for $\nu\to\infty$.
This can be seen by manipulating the conditions (\ref{eq:wbp-symm-disc}) from which we can write
\begin{equation}
	e^{-\nu z} = 2 \frac{(z-1)\nu+1}{2(z-1)\nu+1}.
\end{equation}
In order for this equation to hold, for $\nu\to\infty$ it must be $\nu z(\nu) \to 0 $  and $f(\nu) \to 0 $, which is our statement.
The line of discontinuous transitions never intersects the line of
continuous transitions.
This can be shown by imposing the conditions $\Phi_{f,\nu}(z) = 1$,
$\Phi^{\prime}_{f,\nu}(z) = 0$ and (\ref{eq:wbp-symm-cont}), which
yield $\nu z=0$, {\ie} the intersection cannot occur at finite $\nu$.
Comparing Fig.~\ref{fig:wpp-ph-diag} and Fig.~\ref{fig:wbp-ph-diag},
it transpires that the most remarkable difference between WPP and WBP
in two layers is the occurrence of a discontinuous transition in the latter.
At low values of seed fraction $f$, we observe a decoupling between a
percolating phase driven by the cascade of activations and a smaller
percolating phase driven by the percolation of seeds.

\subsection{Two non-identical {\er} networks}

We apply the above formalism to two uncorrelated {\er} networks with
mean degrees $\mu_1$ and $\mu_2$.
In this case, it so happens that $Z_1$ and $Z_2$ obey identical
equations, so that equation (\ref{eq:Z-wbp-general}) becomes
 
\begin{equation}
Z_1 = Z_2 = pf + p(1-f)(1 - e^{-\mu_1Z_1})(1 - e^{-\mu_2Z_2})
\end{equation}
so we can define $Z \equiv Z_1 = Z_2$, giving
\begin{equation}
Z = pf + p(1-f)(1 - e^{-\mu_1Z})(1 - e^{-\mu_2Z}).
\end{equation}
Similarly, we also find that $X_1 = X_2 \equiv X$, with
\begin{multline}
X = pf(1-e^{-(\mu_1+\mu_2)X}) + p(1-f)[1 -  e^{-(\mu_1+\mu_2)X}\\
-e^{-\mu_1Z}(1-e^{-\mu_2X}) - e^{-\mu_2Z}(1-e^{-\mu_1X})].
\end{multline}
This symmetry, which applies only in the {\er} case, simplifies the
calculations significantly.

Once again we can define rescaled variables $x \equiv X/p$, $z \equiv
Z/p$ and $\nu_1 = p\mu_1$, $\nu_2 = p\mu_2$, giving
\begin{equation}
\Phi_{f,\nu_1,\nu_2}(z) \equiv \frac{1}{z}[f + (1-f)(1-e^{-\nu_1z})(1-e^{-\nu_2z})] = 1
\end{equation}
and
\begin{multline}\label{eq:WBP_x_asym}
x = 1 - e^{-(\nu_1+\nu_2)x} -\\
  (1-f)[e^{-\nu_1z}(1-e^{-\nu_2x})+e^{-\nu_2z}(1-e^{-\nu_1x}) ]\,.
\end{multline}

We look for a continuous appearance of the giant weakly percolating
cluster. From Eq. (\ref{eq:WBP_x_asym}) in the limit of small $x$, we
find the line of continuous transitions given by
\begin{equation}
1-f = \frac{\nu_1+\nu_2-1}{\nu_1e^{-\nu_2z}+\nu_2e^{-\nu_1z}}
\end{equation}
where on this line, $z$ solves
\begin{equation}
1-z = \frac{(\nu_1+\nu_2-1)(e^{-\nu_1z}+e^{-\nu_2z}-e^{-(\nu_1+\nu_2)z})}{\nu_1e^{-\nu_2z}+\nu_2e^{-\nu_1z}}\,.
\end{equation}

\begin{figure}[htb]
\begin{center}
	\includegraphics[width=0.95\columnwidth]{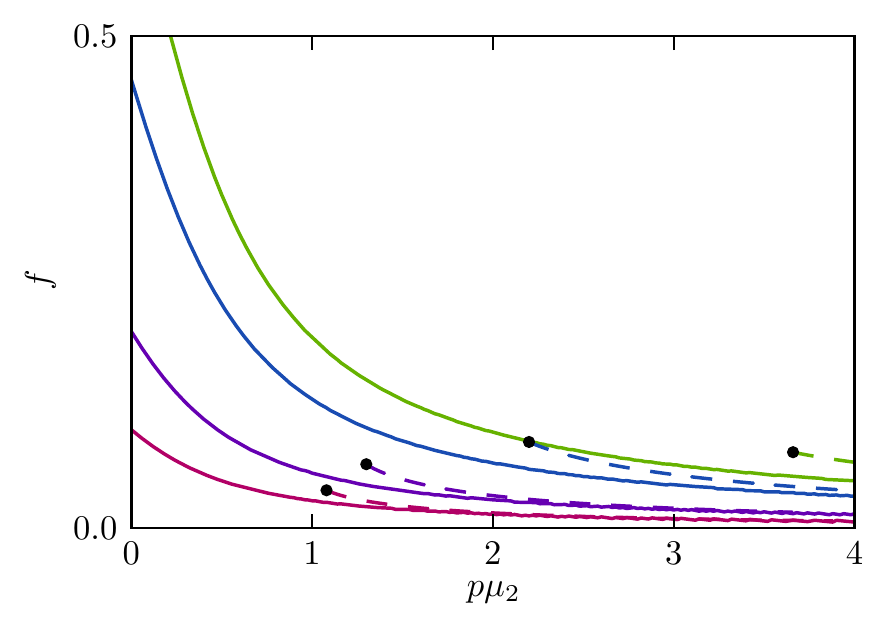}
	\caption{Phase diagram of the WBP model for two uncorrelated
          {\er} networks with mean degree $\mu_1$ and
          $\mu_2$. Horizontal axis is $\nu_2 = p\mu_2$. Each solid
          curve shows the location of the continuous transition for a
          particular value of $\nu_1$, from top to bottom $\nu_1 =
          \{1.5, 2.193, 5, 10\}$.
          Dashed curves show the corresponding location of the
          discontinuous transition (which is always above the
          continuous transition), with circles marking the critical
          end point. Color online.}
	\label{fig:wbp-ph-diag_ERasym}
\end{center}
\end{figure}

The function $\Phi_{f,\nu_1,\nu_2} (z)$ is not always monotonic, so
the phase diagram also contains a line of discontinuous
transitions. This line is defined by the conditions
\begin{equation}
	\left\{%
		\begin{array}{lcr}
			\Phi_{f,\nu_1,\nu_2}(z) &=& 1\\
			\Phi^{\prime}_{f,\nu_1,\nu_2}(z) &=& 0
		\end{array}
	\right.
	\label{eq:wbp-asym-disc}
\end{equation}
The line ends at the critical point defined by these two conditions in
combination with a third condition
\begin{equation}
\Phi^{\prime\prime}_{f,\nu_1,\nu_2}(z) = 0.
\end{equation}
The three conditions can be written
\begin{align}
1\! -\! z &= (1-f)(e^{-\nu_1z}+e^{-\nu_2z}-e^{-(\nu_1+\nu_2)z})\label{eq:WBP_fc_cond1}\\
1 &=
(1\!-\!f)(\nu_1e^{-\nu_1z}+\nu_2e^{-\nu_2z}-(\nu_1+\nu_2)e^{-(\nu_1+\nu_2)z})\label{eq:WBP_fc_cond2}\\
0 &= \nu_1^2e^{-\nu_1z}+\nu_2^2e^{-\nu_2z}-(\nu_1+\nu_2)^2e^{-(\nu_1+\nu_2)z}\label{eq:WBP_fc_cond3}
\end{align}
The phase diagram is plotted in Figure \ref{fig:wbp-ph-diag_ERasym}.
The critical final point occurs generally at small values of $f$,
with the maximum $f_C$ occurring when $\nu_1 = \nu_2 \approx 2.193$
(see figure).
When $\nu_1$ becomes large, numerical solution suggests that  $f_C \to
0$ at a finite value of $\nu_2$.
To check this, we consider
Eqs. (\ref{eq:WBP_fc_cond1})-(\ref{eq:WBP_fc_cond3}) in the large $\nu_1$
limit. Suppose that $z$ remained finite in this limit.  Then
$e^{-\nu_1z} \to 0$, and
Eq. (\ref{eq:WBP_fc_cond3}) implies that $\nu_2 \to 0$ which would
violate Eq. (\ref{eq:WBP_fc_cond2}). We conclude that $z \to 0$ such that
$e^{-\nu_1z}$ remains non-zero.
Under this assumption, and using that $\nu_2/\nu_1 \ll 1$, (\ref{eq:WBP_fc_cond3})
gives us that $\nu_1z \to 2$. Substituting back into (\ref{eq:WBP_fc_cond2})
and using this equation to eliminate $f$ from (\ref{eq:WBP_fc_cond1}), we
find that
\begin{equation}
\nu_2 \to \frac{1}{1 + e^{-2}} \approx 0.88.
\end{equation}
This in turn gives that, indeed, $f_C \to 0$, in the limit $\nu_1 \to
\infty$. This result is confirmed by numerical solution of
Eqs. (\ref{eq:WBP_fc_cond1})-(\ref{eq:WBP_fc_cond3}).

This analysis of the behavior of  two different {\er} networks shows
that the critical point found in the identical case is indeed robust
when we break the symmetry of the two layers.
%
%
%

\section{Critical clusters}
\label{sec:clusters}

To understand the discontinuous transitions which we observe, we now
analyze clusters of critical vertices, through which the avalanches of
damage or activation propagate. Diverging avalanche sizes lead to the
discontinuous transitions.
A critical vertex is a vertex that only just meets the criteria for
inclusion in the percolating cluster (in the case of WPP), or only
just fails to meet the criteria (in the case of WBP).

\subsection{WPP}

\begin{figure}[htb]
\begin{center}
	\includegraphics[width=0.7\columnwidth]{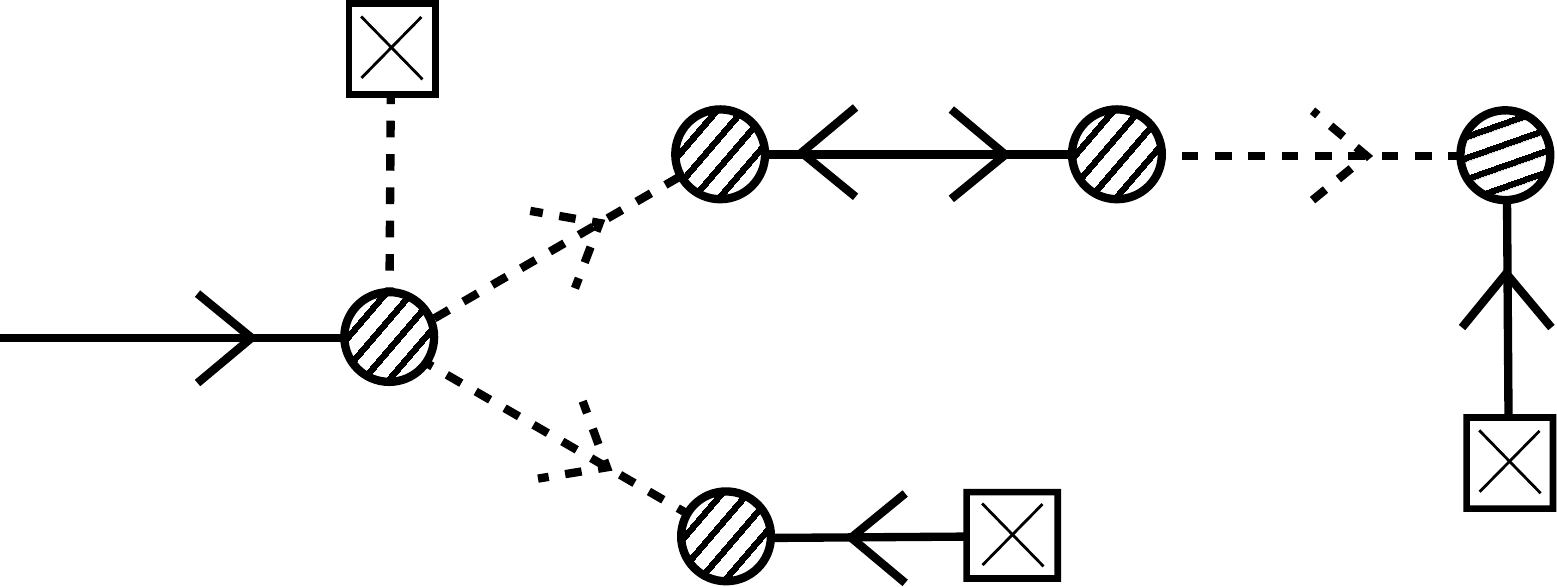}
	\caption{A representation of a cluster of critical vertices in
          WPP. Hatching indicates that vertices are members of the WPP
          percolating cluster. Because critical vertices are in the percolating
          cluster for WPP, a critical vertex may be linked to the
          percolating cluster via another critical vertex. That is,
          external edges of type $Z_i$ are not necessarily
          required. Furthermore, this means that critical dependencies
          can be bi-directional: it is possible for avalanches to
          propagate in either direction along such edges. Note that
          outgoing critical edges must be of the opposite type to the
          incoming one. The boxes containing crosses represent the
          probability $Z_n-R_n$.}
	\label{fig:wpp-avalanche}
\end{center}
\end{figure}

In the case of WPP 
, a
critical vertex is a node that exactly meets the inclusion criterion
. That is, it has exactly one neighbor in the WPP
cluster of type $m$, and at least one of all the other types. We will
call such a vertex a critical vertex of type $m$. A vertex may be
critical with respect to more than one type, if it simultaneously has
exactly one WPP neighbor in more than one layer.
Such a vertex is related to avalanches because it has one (or
possibly more) edge(s) which, if lost, will cause the vertex to be
pruned from the cluster. If, in turn, other outgoing edges of this vertex 
are critical edges for other critical vertices, these vertices
will also be removed. Chains of such connections therefore delineate
the paths of avalanches of spreading damage. An example is shown in
Fig. \ref{fig:wpp-avalanche}. Damage to the node at one end of a
vertex is transmitted along arrowed vertices.

\begin{figure}[htb]
\begin{center}
	\includegraphics[width=0.7\columnwidth]{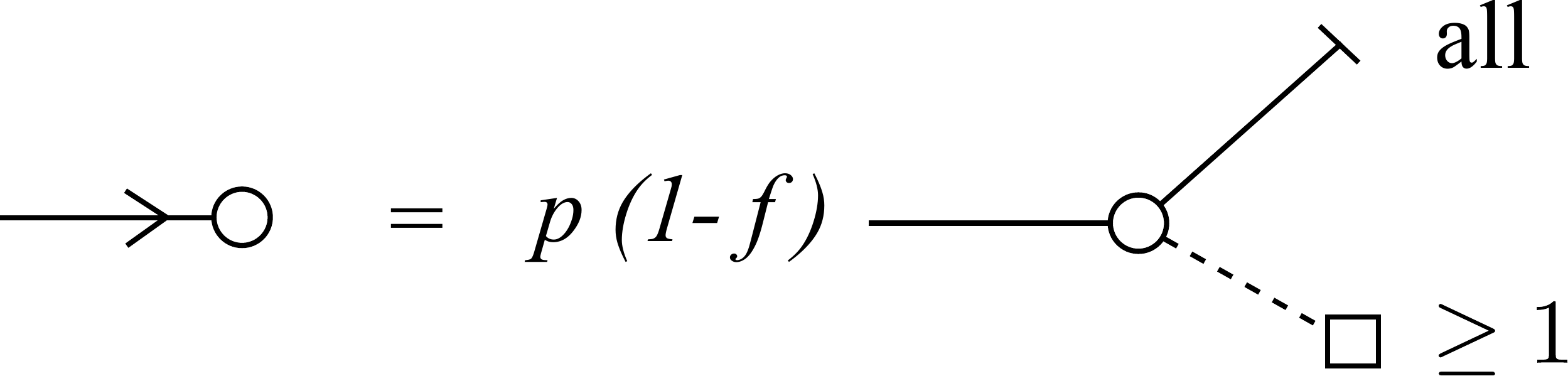}
	\caption{Representation of the probability $R_1$ that, on
          following an edge of type $1$, we encounter a (non-damaged
          and non-seed) vertex that
          has $\geq 1$ child edge of type $2$ leading to a member
          of the percolating cluster and zero of type $1$. The barred
        line represents the probability $1-Z_1$. The arrow in the edge
        is to illustrate the direction of propagation of activation.}
	\label{fig:wbp-R1}
\end{center}
\end{figure}

To examine these avalanches, we define the probability $R_m$, to be
the probability that, on following an edge of type $m$, we encounter a
vulnerable vertex (probability $1-f$), which has not been
removed due to random damage (probability $p$) and has at least one child
edge of each type $n \neq m$ leading to a member of the percolating cluster
(probability $Z_n$), and zero of type $m$. That is
\begin{equation}\label{eq:R1_WPP}
R_m = p(1-f) \sum_{\vec{q}}\frac{q_m P_{\vec{q}}}{\langle q_m\rangle}
(1-Z_m)^{q_m-1} \prod_{\begin{subarray}{c}
        			n=1\\
			n\neq m
      		\end{subarray}}^{M}
\left[ 1 - (1-Z_n)^{q_n}\right].
\end{equation}
This probability, in the case of two layers, is
represented graphically in Figure \ref{fig:wbp-R1}.
We can then define a generating function for the size of the critical
subtree encountered upon following an edge of type $m$ (and hence
resulting activation avalanche should the parent vertex of that edge
be activated) in a recursive way by
\begin{equation}\label{eq:H1_WPP}
H_m(\vec{u}) = Z_m - R_m + u_m F_m[H_1(\vec{u}), H_2(\vec{u}),...,H_M(\vec{u})].
\end{equation}
 The functions $F_m(\vec{x})$ are defined to be
\begin{multline}\label{eq:F1_WPP}
F_m(\vec{x}) = p(1-f) \sum_{\vec{q}}\frac{q_m P_{\vec{q}}}{\langle q_m\rangle}
(1 - Z_m)^{q_m-1}\\
 \prod_{\begin{subarray}{c}
        			n=1\\
			n\neq m
      		\end{subarray}}^{M}
\sum_{l=1}^{q_n} \binom{q_n}{l} (1 - Z_n)^{q_n-l} x_n^{l}.
\end{multline}
Notice that $F_m$ has no dependence on $x_m$.
This method is very similar to that used in \cite{baxter2012}.
A factor $u_m$ appears for every critical edge of type $m$ appearing
in the subtree. The first terms $Z_m-R_m$ give the probability that
zero critical nodes are encountered. The second term, with factor
$u_m$, counts the cases where the first node encountered is a critical
one. This node may have outgoing edges leading to further critical
nodes. These edges are counted by the function $F_m$, and the use of
the generating functions $H_n$ as arguments recursively counts the
size of the critical subtree reached upon following each of these edges.

The mean size of the avalanche caused by the removal of single vertex
is then given by 
\begin{equation}
\sum_{m=1}^M \partial_{u_m}H_m(\vec{1})\,.
\end{equation}
Where $\partial_z$ will be used henceforth to signify the partial
derivative with respect to variable $z$.

Let us first examine the mean avalanche size in the case of two layers.
Taking partial derivatives of Eqs. (\ref{eq:H1_WPP}) and (\ref{eq:F1_WPP}), and after
some rearranging, we arrive at
\begin{equation}
\partial_{u_1} H_1(1,1) = \frac{R_1} {1 -
  \partial_{x_2} F_1(Z_1,Z_2)\partial_{x_1}F_2(Z_1,Z_2) }.
\end{equation}
where we have used that $F_1(Z_1,Z_2) = R_1$ and also that
$H_1(1,1) = Z_1$, and $H_2(1,1) = Z_2$.

Let us define the right-hand side of Eq.~(\ref{eq:Z-wpp-general}) to be
$\Psi_1(Z_1,Z_2)$. 
From Eq.~(\ref{eq:Z-wpp-general}), and comparing with Eq. (\ref{eq:F1_WPP}), the
partial derivatives of $\Psi_1(Z_1,Z_2)$, are
\begin{eqnarray}
\frac{\partial\Psi_1}{\partial Z_1} &=& 0 \nonumber\\
\frac{\partial\Psi_1}{\partial Z_2} &=&  p(1-f)\sum_{q_1,q_2}\frac{
  P_{q_1,q_2}}{\langle q_1\rangle} q_1q_2(1-Z_2)^{q_2-1} \nonumber\\
&& =
\frac{\langle q_2\rangle}{\langle q_1\rangle}\frac{\partial}{\partial_{x_1}}F_2(Z_1,Z_2).
\end{eqnarray}
and similarly for ${\partial\Psi_2}/{\partial Z_1}$ and
${\partial\Psi_2}/{\partial Z_2}$.
Substituting back, we find that
\begin{equation}
\partial_u H_1(1,1) = \frac{R_1}{(\partial \Psi_1/\partial
  Z_2)(\partial \Psi_2/\partial
  Z_1)}.
\end{equation}
The denominator remains finite, and the numerator does not diverge, so
this quantity remains finite everywhere in the 2-layer WPP model.

Following the same procedure in the case of three layers reveals that
\begin{multline}
\partial_{u_1} H_1(1,1) = \\
R_1 \left\{1 - \frac{\partial_2\Psi_1[\partial_1\Psi_2 +
    \partial_1\Psi_3\partial_3\Psi_2]}
{1 - \partial_2\Psi_3\partial_3\Psi_2} - \partial_1\Psi_3\partial_3\Psi_1
\right\}^{-1}\\
 = \frac{R_1}{1 - \frac{d \Psi_1}{d Z_1}}.
\end{multline}
Where for compactness we have written $\partial_m\Psi_n$ for $\partial
\Psi_n /\partial Z_m$.
Now, an alternative form for the condition for the location of the
discontinuous transition is $\frac{d \Psi_1}{d Z_1} = 1$. We see
immediately that this implies that the mean avalanche size diverges at
the critical point.
 In other words the
avalanches diverge in size as the discontinuous hybrid transition
approaches, just as the susceptibility does for an ordinary
second-order transition. 

\subsection{WBP}

\begin{figure}[htb]
\begin{center}
	\includegraphics[width=0.7\columnwidth]{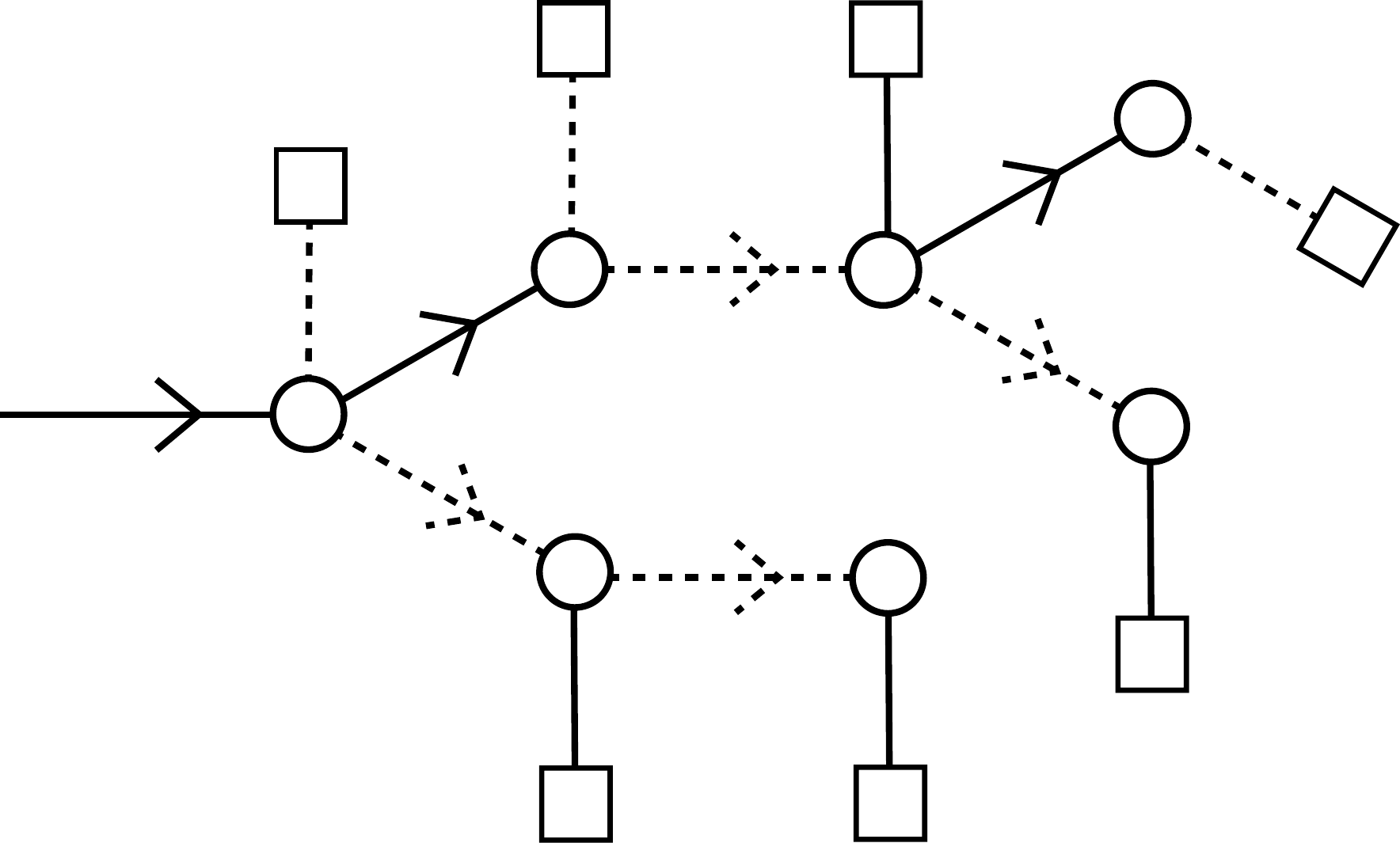}
	\caption{An example of a critical cluster in WBP. Avalanches
          of activation propagate through the cluster following the
          arrowed edges. If an upstream vertex is activated, all
          downstream critical vertices will in turn be activated. Note
        that, unlike for WPP,  in WBP it is not possible for an edge to be arrowed in both directions. Activation can only ever propagate in one
        direction along a given edge. Also note
          that in the WBP case outgoing critical
          edges must be of the same type as the incoming one.}
	\label{fig:wbp-avalanche}
\end{center}
\end{figure}

In the case of WBP, a
critical vertex is one that fails the inclusion criterion for a single type of
edge. That is, it has exactly zero neighbors in the WBP cluster of
type $m$, for example, and at least one  of every other type.
Such a vertex is related to avalanches because, if it gains a single
connection of type $m$ to an active node, it will itself join the active
WBP cluster. If, in turn, other outgoing edges of this vertex 
are the critical edges for other critical vertices, these vertices
will also become active. Chains of such connections therefore delineate
the paths of avalanches of spreading activation. An example of a small
critical cluster is shown in Figure \ref{fig:wbp-avalanche}.

To examine these avalanches, we define the probability $R_m$, to be
the probability that, on following an edge of type $m$, we encounter a
vertex which is not a seed vertex (probability $1-f$), has not been
removed due to random damage (probability $p$) has at least one child
edge of all other types $n\neq m$ leading to members of the percolating cluster
(probability $Z_n$), and zero of type $m$. That is
\begin{equation}\label{eq:R1_WBP}
R_m = p(1-f) \sum_{\vec{q}}\frac{q_m P_{\vec{q}}}{\langle q_m\rangle}
(1-Z_m)^{q_m-1}  \prod_{\begin{subarray}{c}
        			n=1\\
			n\neq m
      		\end{subarray}}^{M}
\left[ 1 - (1-Z_n)^{q_n}\right]
\end{equation}
Note that this is identical to
(\ref{eq:R1_WPP}), but the following argument is different.

Because critical vertices are outside the WBP cluster, the
probabilities $Z_m$ and $R_m$ are mutually
exclusive. This means that, upon following an edge of type $m$, there
are 3 non-overlapping possibilities: we encounter a percolating vertex
(probability $Z_m$) we encounter a critical vertex (probability $R_m$)
or we encounter neither (probability $1-Z_m-R_m$).
We can then define a generating function for the size of the critical
subtree encountered upon following an edge of type $m$ (and hence
resulting activation avalanche should the parent vertex of that edge
be activated) in a recursive way by
\begin{equation}\label{eq:H1_WBP}
H_m(\vec{u}) = 1 - Z_m - R_m + u_m F_m[H_1(\vec{u}), H_2(\vec{u}),...,H_M(\vec{u})]\,.
\end{equation}
The functions $F_m(\vec{x})$ are defined to be
\begin{equation}\label{eq:F1_WBP}
F_m(x,y) = p(1-f) \sum_{\vec{q}}\frac{q_m P_{\vec{q}}}{\langle q_m\rangle}
x_m^{q_m-1} \prod_{\begin{subarray}{c}
        			n=1\\
			n\neq m
      		\end{subarray}}^{M}
 \sum_{l=1}^{q_n} \binom{q_n}{l} Z_n^l x_n^{q_n-l}.
\end{equation}
Note that $F_m(1-Z_1,1-Z_2,...,1-Z_M) = R_m$ and $H_m(\vec{1}) = 1-Z_m$.

The mean size of the avalanche caused by the activation of a single
vertex is again given by 
\begin{equation}
\sum_{m=1}^M \partial_{u_m}H_m(\vec{1}).
\end{equation}

Let us consider the case of WBP in a 2-layer multiplex.
Taking partial derivatives of (\ref{eq:H1_WBP}) and (\ref{eq:F1_WBP}) and
after some rearranging, we find
\begin{widetext}
\begin{equation}
\partial_{u_1} H_1(1,1) = \frac{R_1\left[1 - \partial_{x_2}F_2(1\!-\!Z_1,1\!-\!Z_2)
    \right]} {\left[1 - \partial_{x_1}F_1(1\!-\!Z_1,1\!-\!Z_2)
    \right]\left[1 - \partial_{x_2}F_2(1\!-\!Z_1,1\!-\!Z_2)
    \right] - \partial_{x_2}F_1(1\!-\!Z_1,1\!-\!Z_2)
  \partial_{x_1}F_2(1\!-\!Z_1,1\!-\!Z_2) },
\end{equation}
\end{widetext}
where we have used that $F_1(1\!-\!Z_1,1\!-\!Z_2) = R_1$ and also that
$H_1(1,1) = 1- Z_1$, and $H_2(1,1) = 1 - Z_2$.

Let us define the right-hand side of Eq.~(\ref{eq:Z-wbp-general}), in
the two layer case, to be
$\Psi_n(Z_1,Z_2)$. Then
\begin{align}
\frac{\partial \Psi_1}{\partial Z_1} &= \nonumber \\
p(1-f)&
\sum_{q_1,q_2}\frac{ P_{q_1,q_2}}{\langle q_1\rangle}q_1(q_1-1)
(1-Z_1)^{q_1-2} [1 - (1-Z_2)^{q_2}]\nonumber\\
 &= \partial_{x_1} F_1(1-Z_1,1-Z_2)
\end{align}
and
\begin{align}
\frac{\partial \Psi_1}{\partial Z_2} &= \nonumber\\
p(1-f) &
\sum_{q_1,q_2}\frac{ P_{q_1,q_2}}{\langle q_1\rangle}q_1q_2
(1-Z_2)^{q_2-1} [1 - (1-Z_1)^{q_1-1}]\nonumber\\
&= \frac{\langle q_2\rangle}{\langle q_1\rangle} \partial_{x_1} F_2(1-Z_1,1-Z_2)
\end{align}
and a similar procedure is followed for $\Psi_2$.
This means that the equation for $\partial_{u_1}H_1(1,1)$ can be written
\begin{equation}
\partial_{u_1}H_1(1,1) = \frac{R_1 [1 - \partial \Psi_2 /\partial Z_2]}
        {\det[{\bf J}-{\bf I}]}\,.
\end{equation}
where the Jacobian matrix ${\bf J}$ has elements $J_{ij} = \partial
\Psi_i/Z_j$, and ${\bf I}$ is the identity matrix.
The condition $\frac{d \Psi_1}{d Z_1} = 1$ for the location of the
discontinuity in $Z_1$ (and $Z_2$) can be rewritten
\begin{equation}
\det[{\bf J}-{\bf I}] = 0
\end{equation}
meaning that $\partial_{u_1}H_1(1,1)$ diverges, and hence the mean
avalanche size, diverges precisely at the critical point.
A similar analysis can be performed for three (and in principle more) layers.

\section{Conclusions}
\label{sec:conclusions}

In this paper we have introduced weak bootstrap percolation (WBP) and weak pruning
percolation (WPP) in multiplex networks. These are natural extensions
of percolation on simplex networks, and are somewhat analogous to
bootstrap percolation and the $k$-core pruning algorithm on simplex
networks. 
We have shown that, unlike the case of a single layer, these two models are distinct and give origin to different critical behaviors. 
We further introduced the concept of invulnerable nodes in
multiplex percolation, and shown their effect on the critical
transitions associated with the emergence of a giant percolating
cluster.
We have explicitly calculated the critical phenomena characterizing multiplex  networks made of {\er} networks on each layer.
The WBP model includes both continuous and discontinuous hybrid transitions
for two or more layers, while the WPP model has only continuous
transitions in 2 layers, but a discontinuous hybrid transition appears
when there are 3 or more layers. The discontinuous transition in the
3-layer WPP disappears at the same critical point as that in 2-layer
WBP. A cursory examination of the relevant equations reveals why there
should be such a connection between the 2-layer WBP and 3-layer
WPP. In $M$-layer WPP, the probability $Z_n$, that an edge of type $n$
leads to a member of the percolating cluster, requires that the node
reached have connections of the other $M-1$ types to percolating
nodes. Compare this with $(M-1)$-layer WBP, where the criterion is
again that the node reached have $M-1$ connections (which in this case
is all types of links).

In a broader context, the present $3$-layer WPP model and $2$-layer
WBP model show phase diagrams similar to that seen in the (1,3)
heterogeneous $k$-core model and (1,2) bootstrap percolation models
respectively \cite{baxter2011}. There are no tricritical points in these
systems, but it was shown in \cite{cellai2011} that a tricritical
point does appear in the (2,3) heterogeneous $k$-core. Noting that a
similar tricritical point appears in partial interdependence models \cite{Gao2011,son2012}, it seems natural to consider that
WPP and the partial interdependence model, as specific cases of a
broader class of mixed-rule multiplex percolation models.
After this paper was submitted, another work \cite{Min2014} appeared,
where two very similar models have been investigated in the context of
infrastructure management.

The WPP model, besides filling a gap in our understanding of percolation on multilayer networks, also provides an interesting diagnostics to evaluate the presence of a missing layer.
The absence of critical points and discontinuous transitions in the 2-layer case, in fact, is remarkable as it qualitatively differs from the 3-layer case, and it might be used to determine the layer structure of multiplex networks with limited information.
A similar qualitative behavior has been recently observed in the case of classical percolation with edge overlap \cite{cellai2013c}.

The WBP model constitutes the simplest activation process which may occur on a multiplex.
Differently from other more complex models, here we provide a simple analytical method which allows not only exact calculation of the critical behavior in locally tree-like networks, but also the calculation of critical exponents and  the characterization of critical clusters.
Moreover, the WBP model has potential  applications in infrastructure recovery and information security.
Wireless sensor networks require key distribution schemes able to guarantee overall secure communication even in the case where some sensors are compromised by an external attack (see eg. \cite{chen2011b} for a review).
The first key distribution scheme specifically designed for sensor networks was introduced  by Eschenauer and Gligor \cite{eschenauer2002}.
This scheme is based on assigning to each sensor a random set of keys taken from a large key pool.
Sensors sharing the same key can communicate directly.
As more secure developments of this scheme, several protocols have been proposed.
One of them prescribes a distribution scheme where each node shares a unique pairwise key with with  each of $O(\sqrt{N})$ other nodes in the network \cite{chen2005}.
This implies that a sensor $A$ may need an intermediary sensor $C$ to establish communication with another sensor $B$.
Let us consider now a network where different edge types correspond to a key shared among two sensors.
If we define a sensor to be compromised when all its keys have been captured by an intruder, then we can view this problem as a bootstrap process, where either a cascade of compromised nodes occurs, or the attack remains confined to a few nodes.

In the case of infrastructure, the same  interaction between
infrastructure layers that can lead to dramatic collapse can lead to
dramatic recovery. After a large disruption,  the
re-activation of a small number of nodes may allow further nodes which
have lost one of their essential dependencies to resume
functioning. Once a certain threshold is reached, this may lead to
dramatic gains in the functioning of the entire system \cite{Duenas2007}.

Finally, our paper shows that extensions of simple traditional models to multiplex networks generate a wealth of new possibilities, both in terms of model definitions and of new critical behaviors, with implications mostly to be understood.
%

We acknowledge useful interactions with  James Gleeson and Adrian Perrig.
This work has been partially funded by Science Foundation Ireland,
grant 11/PI/1026; the FET-Proactive project PLEXMATH
(FP7-ICT-2011-8; grant number 317614), FET IP Project
MULTIPLEX 317532; by the Portuguese Science and Technology Foundation (FCT)
Project PEst-C/CTM/LA0025/2011, and Grant No. SFRH/BPD/74040/2010.


\bibliography{library,network}

\begin{thebibliography}{33}%
\makeatletter
\providecommand \@ifxundefined [1]{%
 \@ifx{#1\undefined}
}%
\providecommand \@ifnum [1]{%
 \ifnum #1\expandafter \@firstoftwo
 \else \expandafter \@secondoftwo
 \fi
}%
\providecommand \@ifx [1]{%
 \ifx #1\expandafter \@firstoftwo
 \else \expandafter \@secondoftwo
 \fi
}%
\providecommand \natexlab [1]{#1}%
\providecommand \enquote  [1]{``#1''}%
\providecommand \bibnamefont  [1]{#1}%
\providecommand \bibfnamefont [1]{#1}%
\providecommand \citenamefont [1]{#1}%
\providecommand \href@noop [0]{\@secondoftwo}%
\providecommand \href [0]{\begingroup \@sanitize@url \@href}%
\providecommand \@href[1]{\@@startlink{#1}\@@href}%
\providecommand \@@href[1]{\endgroup#1\@@endlink}%
\providecommand \@sanitize@url [0]{\catcode `\\12\catcode `\$12\catcode
  `\&12\catcode `\#12\catcode `\^12\catcode `\_12\catcode `\%12\relax}%
\providecommand \@@startlink[1]{}%
\providecommand \@@endlink[0]{}%
\providecommand \url  [0]{\begingroup\@sanitize@url \@url }%
\providecommand \@url [1]{\endgroup\@href {#1}{\urlprefix }}%
\providecommand \urlprefix  [0]{URL }%
\providecommand \Eprint [0]{\href }%
\providecommand \doibase [0]{http://dx.doi.org/}%
\providecommand \selectlanguage [0]{\@gobble}%
\providecommand \bibinfo  [0]{\@secondoftwo}%
\providecommand \bibfield  [0]{\@secondoftwo}%
\providecommand \translation [1]{[#1]}%
\providecommand \BibitemOpen [0]{}%
\providecommand \bibitemStop [0]{}%
\providecommand \bibitemNoStop [0]{.\EOS\space}%
\providecommand \EOS [0]{\spacefactor3000\relax}%
\providecommand \BibitemShut  [1]{\csname bibitem#1\endcsname}%
\let\auto@bib@innerbib\@empty
\bibitem [{\citenamefont {Caccioli}\ \emph {et~al.}(2012)\citenamefont
  {Caccioli}, \citenamefont {Shrestha}, \citenamefont {Moore},\ and\
  \citenamefont {Farmer}}]{caccioli2012}%
  \BibitemOpen
  \bibfield  {author} {\bibinfo {author} {\bibfnamefont {F.}~\bibnamefont
  {Caccioli}}, \bibinfo {author} {\bibfnamefont {M.}~\bibnamefont {Shrestha}},
  \bibinfo {author} {\bibfnamefont {C.}~\bibnamefont {Moore}}, \ and\ \bibinfo
  {author} {\bibfnamefont {J.~D.}\ \bibnamefont {Farmer}},\ }\href
  {http://arxiv.org/abs/1210.5987} {\enquote {\bibinfo {title} {Stability
  analysis of financial contagion due to overlapping portfolios},}\ } (\bibinfo
  {year} {2012}),\ \Eprint {http://arxiv.org/abs/1210.5987} {arXiv:1210.5987}
  \BibitemShut {NoStop}%
\bibitem [{\citenamefont {Huang}\ \emph {et~al.}(2013)\citenamefont {Huang},
  \citenamefont {Vodenska}, \citenamefont {Havlin},\ and\ \citenamefont
  {Stanley}}]{Huang2013}%
  \BibitemOpen
  \bibfield  {author} {\bibinfo {author} {\bibfnamefont {X.}~\bibnamefont
  {Huang}}, \bibinfo {author} {\bibfnamefont {I.}~\bibnamefont {Vodenska}},
  \bibinfo {author} {\bibfnamefont {S.}~\bibnamefont {Havlin}}, \ and\ \bibinfo
  {author} {\bibfnamefont {H.~E.}\ \bibnamefont {Stanley}},\ }\href@noop {}
  {\bibfield  {journal} {\bibinfo  {journal} {Scientific reports}\ }\textbf
  {\bibinfo {volume} {3}},\ \bibinfo {pages} {1219} (\bibinfo {year}
  {2013})}\BibitemShut {NoStop}%
\bibitem [{\citenamefont {Pocock}\ \emph {et~al.}(2012)\citenamefont {Pocock},
  \citenamefont {Evans},\ and\ \citenamefont {Memmott}}]{Pocock2012}%
  \BibitemOpen
  \bibfield  {author} {\bibinfo {author} {\bibfnamefont {M.~J.~O.}\
  \bibnamefont {Pocock}}, \bibinfo {author} {\bibfnamefont {D.~M.}\
  \bibnamefont {Evans}}, \ and\ \bibinfo {author} {\bibfnamefont
  {J.}~\bibnamefont {Memmott}},\ }\href@noop {} {\bibfield  {journal} {\bibinfo
   {journal} {Science}\ }\textbf {\bibinfo {volume} {335}},\ \bibinfo {pages}
  {973} (\bibinfo {year} {2012})}\BibitemShut {NoStop}%
\bibitem [{\citenamefont {Rinaldi}\ \emph {et~al.}(2001)\citenamefont
  {Rinaldi}, \citenamefont {Peerenboom},\ and\ \citenamefont
  {Kelly}}]{Rinaldi2001}%
  \BibitemOpen
  \bibfield  {author} {\bibinfo {author} {\bibfnamefont {S.~M.}\ \bibnamefont
  {Rinaldi}}, \bibinfo {author} {\bibfnamefont {J.~P.}\ \bibnamefont
  {Peerenboom}}, \ and\ \bibinfo {author} {\bibfnamefont {T.~K.}\ \bibnamefont
  {Kelly}},\ }\href@noop {} {\bibfield  {journal} {\bibinfo  {journal} {IEEE
  Control Syst. Mag.}\ }\textbf {\bibinfo {volume} {21}},\ \bibinfo {pages}
  {11} (\bibinfo {year} {2001})}\BibitemShut {NoStop}%
\bibitem [{\citenamefont {Leicht}\ and\ \citenamefont
  {D'Souza}(2009)}]{leicht2009}%
  \BibitemOpen
  \bibfield  {author} {\bibinfo {author} {\bibfnamefont {E.~A.}\ \bibnamefont
  {Leicht}}\ and\ \bibinfo {author} {\bibfnamefont {R.~M.}\ \bibnamefont
  {D'Souza}},\ }\href {http://arxiv.org/abs/0907.0894} {\enquote {\bibinfo
  {title} {Percolation on interacting networks},}\ } (\bibinfo {year} {2009}),\
  \Eprint {http://arxiv.org/abs/0907.0894} {arXiv:0907.0894} \BibitemShut
  {NoStop}%
\bibitem [{\citenamefont {Kivel\"{a}}\ \emph {et~al.}(2013)\citenamefont
  {Kivel\"{a}}, \citenamefont {Arenas}, \citenamefont {Barthelemy},
  \citenamefont {Gleeson}, \citenamefont {Moreno},\ and\ \citenamefont
  {Porter}}]{kivela2013}%
  \BibitemOpen
  \bibfield  {author} {\bibinfo {author} {\bibfnamefont {M.}~\bibnamefont
  {Kivel\"{a}}}, \bibinfo {author} {\bibfnamefont {A.}~\bibnamefont {Arenas}},
  \bibinfo {author} {\bibfnamefont {M.}~\bibnamefont {Barthelemy}}, \bibinfo
  {author} {\bibfnamefont {J.~P.}\ \bibnamefont {Gleeson}}, \bibinfo {author}
  {\bibfnamefont {Y.}~\bibnamefont {Moreno}}, \ and\ \bibinfo {author}
  {\bibfnamefont {M.~A.}\ \bibnamefont {Porter}},\ }\href
  {http://arxiv.org/abs/1309.7233} {\enquote {\bibinfo {title} {Multilayer
  networks},}\ } (\bibinfo {year} {2013}),\ \Eprint
  {http://arxiv.org/abs/1309.7233} {arXiv:1309.7233} \BibitemShut {NoStop}%
\bibitem [{\citenamefont {Buldyrev}\ \emph {et~al.}(2010)\citenamefont
  {Buldyrev}, \citenamefont {Parshani}, \citenamefont {Stanley},\ and\
  \citenamefont {Havlin}}]{Buldyrev2010}%
  \BibitemOpen
  \bibfield  {author} {\bibinfo {author} {\bibfnamefont {S.~V.}\ \bibnamefont
  {Buldyrev}}, \bibinfo {author} {\bibfnamefont {R.}~\bibnamefont {Parshani}},
  \bibinfo {author} {\bibfnamefont {G.~P. H.~E.}\ \bibnamefont {Stanley}}, \
  and\ \bibinfo {author} {\bibfnamefont {S.}~\bibnamefont {Havlin}},\
  }\href@noop {} {\bibfield  {journal} {\bibinfo  {journal} {Nature}\ }\textbf
  {\bibinfo {volume} {464}},\ \bibinfo {pages} {08932} (\bibinfo {year}
  {2010})}\BibitemShut {NoStop}%
\bibitem [{\citenamefont {De~Domenico}\ \emph {et~al.}(2013)\citenamefont
  {De~Domenico}, \citenamefont {Sol\'{e}-Ribalta}, \citenamefont {Cozzo},
  \citenamefont {Kivel\"{a}}, \citenamefont {Moreno}, \citenamefont {Porter},
  \citenamefont {G\'{o}mez},\ and\ \citenamefont {Arenas}}]{dedomenico2013}%
  \BibitemOpen
  \bibfield  {author} {\bibinfo {author} {\bibfnamefont {M.}~\bibnamefont
  {De~Domenico}}, \bibinfo {author} {\bibfnamefont {A.}~\bibnamefont
  {Sol\'{e}-Ribalta}}, \bibinfo {author} {\bibfnamefont {E.}~\bibnamefont
  {Cozzo}}, \bibinfo {author} {\bibfnamefont {M.}~\bibnamefont {Kivel\"{a}}},
  \bibinfo {author} {\bibfnamefont {Y.}~\bibnamefont {Moreno}}, \bibinfo
  {author} {\bibfnamefont {M.~A.}\ \bibnamefont {Porter}}, \bibinfo {author}
  {\bibfnamefont {S.}~\bibnamefont {G\'{o}mez}}, \ and\ \bibinfo {author}
  {\bibfnamefont {A.}~\bibnamefont {Arenas}},\ }\href {\doibase
  10.1103/PhysRevX.3.041022} {\bibfield  {journal} {\bibinfo  {journal}
  {Physical Review X}\ }\textbf {\bibinfo {volume} {3}},\ \bibinfo {pages}
  {041022+} (\bibinfo {year} {2013})},\ \Eprint
  {http://arxiv.org/abs/1307.4977} {arXiv:1307.4977} \BibitemShut {NoStop}%
\bibitem [{\citenamefont {Bianconi}(2013)}]{bianconi2013}%
  \BibitemOpen
  \bibfield  {author} {\bibinfo {author} {\bibfnamefont {G.}~\bibnamefont
  {Bianconi}},\ }\href {\doibase 10.1103/physreve.87.062806} {\bibfield
  {journal} {\bibinfo  {journal} {Physical Review E}\ }\textbf {\bibinfo
  {volume} {87}},\ \bibinfo {pages} {062806+} (\bibinfo {year}
  {2013})}\BibitemShut {NoStop}%
\bibitem [{\citenamefont {Son}\ \emph {et~al.}(2011)\citenamefont {Son},
  \citenamefont {Grassberger},\ and\ \citenamefont {Paczuski}}]{Son2011}%
  \BibitemOpen
  \bibfield  {author} {\bibinfo {author} {\bibfnamefont {S.-W.}\ \bibnamefont
  {Son}}, \bibinfo {author} {\bibfnamefont {P.}~\bibnamefont {Grassberger}}, \
  and\ \bibinfo {author} {\bibfnamefont {M.}~\bibnamefont {Paczuski}},\
  }\href@noop {} {\bibfield  {journal} {\bibinfo  {journal} {Phys. Rev. Lett.}\
  }\textbf {\bibinfo {volume} {107}},\ \bibinfo {pages} {195702} (\bibinfo
  {year} {2011})}\BibitemShut {NoStop}%
\bibitem [{\citenamefont {Buldyrev}\ \emph {et~al.}(2011)\citenamefont
  {Buldyrev}, \citenamefont {Shere},\ and\ \citenamefont
  {Cwilich}}]{Buldyrev2011}%
  \BibitemOpen
  \bibfield  {author} {\bibinfo {author} {\bibfnamefont {S.~V.}\ \bibnamefont
  {Buldyrev}}, \bibinfo {author} {\bibfnamefont {N.~W.}\ \bibnamefont {Shere}},
  \ and\ \bibinfo {author} {\bibfnamefont {G.~A.}\ \bibnamefont {Cwilich}},\
  }\href@noop {} {\bibfield  {journal} {\bibinfo  {journal} {Phys. Rev. E}\
  }\textbf {\bibinfo {volume} {83}},\ \bibinfo {pages} {016112} (\bibinfo
  {year} {2011})}\BibitemShut {NoStop}%
\bibitem [{\citenamefont {Baxter}\ \emph {et~al.}(2012)\citenamefont {Baxter},
  \citenamefont {Dorogovtsev}, \citenamefont {Goltsev},\ and\ \citenamefont
  {Mendes}}]{baxter2012}%
  \BibitemOpen
  \bibfield  {author} {\bibinfo {author} {\bibfnamefont {G.~J.}\ \bibnamefont
  {Baxter}}, \bibinfo {author} {\bibfnamefont {S.~N.}\ \bibnamefont
  {Dorogovtsev}}, \bibinfo {author} {\bibfnamefont {A.~V.}\ \bibnamefont
  {Goltsev}}, \ and\ \bibinfo {author} {\bibfnamefont {J.~F.~F.}\ \bibnamefont
  {Mendes}},\ }\href {\doibase 10.1103/PhysRevLett.109.248701} {\bibfield
  {journal} {\bibinfo  {journal} {Physical Review Letters}\ }\textbf {\bibinfo
  {volume} {109}} (\bibinfo {year} {2012}),\ 10.1103/PhysRevLett.109.248701},\
  \Eprint {http://arxiv.org/abs/1207.0448} {arXiv:1207.0448} \BibitemShut
  {NoStop}%
\bibitem [{\citenamefont {Due{\~n}as}\ \emph {et~al.}(2007)\citenamefont
  {Due{\~n}as}, \citenamefont {Cragin},\ and\ \citenamefont
  {Goodno}}]{Duenas2007}%
  \BibitemOpen
  \bibfield  {author} {\bibinfo {author} {\bibfnamefont {L.}~\bibnamefont
  {Due{\~n}as}}, \bibinfo {author} {\bibfnamefont {J.~I.}\ \bibnamefont
  {Cragin}}, \ and\ \bibinfo {author} {\bibfnamefont {B.~J.}\ \bibnamefont
  {Goodno}},\ }\href@noop {} {\bibfield  {journal} {\bibinfo  {journal}
  {Earthquake Eng. Struct. Dynam.}\ }\textbf {\bibinfo {volume} {36}},\
  \bibinfo {pages} {285} (\bibinfo {year} {2007})}\BibitemShut {NoStop}%
\bibitem [{\citenamefont {Lee}\ \emph {et~al.}(2007)\citenamefont {Lee},
  \citenamefont {Mitchell},\ and\ \citenamefont {Wallace}}]{Lee2007}%
  \BibitemOpen
  \bibfield  {author} {\bibinfo {author} {\bibfnamefont {E.~E.}\ \bibnamefont
  {Lee}}, \bibinfo {author} {\bibfnamefont {J.~E.}\ \bibnamefont {Mitchell}}, \
  and\ \bibinfo {author} {\bibfnamefont {W.~A.}\ \bibnamefont {Wallace}},\
  }\href@noop {} {\bibfield  {journal} {\bibinfo  {journal} {IEEE Trans. Syst.,
  Man, Cybern., Syst. C}\ }\textbf {\bibinfo {volume} {37}},\ \bibinfo {pages}
  {1303} (\bibinfo {year} {2007})}\BibitemShut {NoStop}%
\bibitem [{\citenamefont {Parshani}\ \emph {et~al.}(2010)\citenamefont
  {Parshani}, \citenamefont {Buldyrev},\ and\ \citenamefont
  {Havlin}}]{Parshani2010}%
  \BibitemOpen
  \bibfield  {author} {\bibinfo {author} {\bibfnamefont {R.}~\bibnamefont
  {Parshani}}, \bibinfo {author} {\bibfnamefont {S.~V.}\ \bibnamefont
  {Buldyrev}}, \ and\ \bibinfo {author} {\bibfnamefont {S.}~\bibnamefont
  {Havlin}},\ }\href {\doibase 10.1103/PhysRevLett.105.048701} {\bibfield
  {journal} {\bibinfo  {journal} {Phys. Rev. Lett.}\ }\textbf {\bibinfo
  {volume} {105}},\ \bibinfo {pages} {048701} (\bibinfo {year}
  {2010})}\BibitemShut {NoStop}%
\bibitem [{\citenamefont {Hu}\ \emph {et~al.}(2013)\citenamefont {Hu},
  \citenamefont {Zhou}, \citenamefont {Zhang}, \citenamefont {Han},
  \citenamefont {Rozenblat},\ and\ \citenamefont {Havlin}}]{hu2013}%
  \BibitemOpen
  \bibfield  {author} {\bibinfo {author} {\bibfnamefont {Y.}~\bibnamefont
  {Hu}}, \bibinfo {author} {\bibfnamefont {D.}~\bibnamefont {Zhou}}, \bibinfo
  {author} {\bibfnamefont {R.}~\bibnamefont {Zhang}}, \bibinfo {author}
  {\bibfnamefont {Z.}~\bibnamefont {Han}}, \bibinfo {author} {\bibfnamefont
  {C.}~\bibnamefont {Rozenblat}}, \ and\ \bibinfo {author} {\bibfnamefont
  {S.}~\bibnamefont {Havlin}},\ }\href {\doibase 10.1103/PhysRevE.88.052805}
  {\bibfield  {journal} {\bibinfo  {journal} {Physical Review E}\ }\textbf
  {\bibinfo {volume} {88}},\ \bibinfo {pages} {052805} (\bibinfo {year}
  {2013})},\ \Eprint {http://arxiv.org/abs/1308.1862} {arXiv:1308.1862}
  \BibitemShut {NoStop}%
\bibitem [{\citenamefont {Cellai}\ \emph
  {et~al.}(2013{\natexlab{a}})\citenamefont {Cellai}, \citenamefont
  {L\'{o}pez}, \citenamefont {Zhou}, \citenamefont {Gleeson},\ and\
  \citenamefont {Bianconi}}]{cellai2013c}%
  \BibitemOpen
  \bibfield  {author} {\bibinfo {author} {\bibfnamefont {D.}~\bibnamefont
  {Cellai}}, \bibinfo {author} {\bibfnamefont {E.}~\bibnamefont {L\'{o}pez}},
  \bibinfo {author} {\bibfnamefont {J.}~\bibnamefont {Zhou}}, \bibinfo {author}
  {\bibfnamefont {J.~P.}\ \bibnamefont {Gleeson}}, \ and\ \bibinfo {author}
  {\bibfnamefont {G.}~\bibnamefont {Bianconi}},\ }\href {\doibase
  10.1103/PhysRevE.88.052811} {\bibfield  {journal} {\bibinfo  {journal}
  {Physical Review E}\ }\textbf {\bibinfo {volume} {88}},\ \bibinfo {pages}
  {052811} (\bibinfo {year} {2013}{\natexlab{a}})},\ \Eprint
  {http://arxiv.org/abs/1307.6359} {arXiv:1307.6359} \BibitemShut {NoStop}%
\bibitem [{\citenamefont {Gao}\ \emph {et~al.}(2011)\citenamefont {Gao},
  \citenamefont {Buldyrev}, \citenamefont {Havlin},\ and\ \citenamefont
  {Stanley}}]{Gao2011}%
  \BibitemOpen
  \bibfield  {author} {\bibinfo {author} {\bibfnamefont {J.}~\bibnamefont
  {Gao}}, \bibinfo {author} {\bibfnamefont {S.~V.}\ \bibnamefont {Buldyrev}},
  \bibinfo {author} {\bibfnamefont {S.}~\bibnamefont {Havlin}}, \ and\ \bibinfo
  {author} {\bibfnamefont {H.~E.}\ \bibnamefont {Stanley}},\ }\href@noop {}
  {\bibfield  {journal} {\bibinfo  {journal} {Phys. Rev. Lett.}\ }\textbf
  {\bibinfo {volume} {107}},\ \bibinfo {pages} {195701} (\bibinfo {year}
  {2011})}\BibitemShut {NoStop}%
\bibitem [{\citenamefont {Son}\ \emph {et~al.}(2012)\citenamefont {Son},
  \citenamefont {Bizhani}, \citenamefont {Christensen}, \citenamefont
  {Grassberger},\ and\ \citenamefont {Paczuski}}]{son2012}%
  \BibitemOpen
  \bibfield  {author} {\bibinfo {author} {\bibfnamefont {S.-W.}\ \bibnamefont
  {Son}}, \bibinfo {author} {\bibfnamefont {G.}~\bibnamefont {Bizhani}},
  \bibinfo {author} {\bibfnamefont {C.}~\bibnamefont {Christensen}}, \bibinfo
  {author} {\bibfnamefont {P.}~\bibnamefont {Grassberger}}, \ and\ \bibinfo
  {author} {\bibfnamefont {M.}~\bibnamefont {Paczuski}},\ }\href {\doibase
  10.1209/0295-5075/97/16006} {\bibfield  {journal} {\bibinfo  {journal} {EPL
  (Europhysics Letters)}\ }\textbf {\bibinfo {volume} {97}},\ \bibinfo {pages}
  {16006+} (\bibinfo {year} {2012})},\ \Eprint {http://arxiv.org/abs/1109.4447}
  {arXiv:1109.4447} \BibitemShut {NoStop}%
\bibitem [{\citenamefont {Min}\ \emph {et~al.}(2013)\citenamefont {Min},
  \citenamefont {Yi}, \citenamefont {Lee},\ and\ \citenamefont
  {Goh}}]{min2013}%
  \BibitemOpen
  \bibfield  {author} {\bibinfo {author} {\bibfnamefont {B.}~\bibnamefont
  {Min}}, \bibinfo {author} {\bibfnamefont {S.~D.}\ \bibnamefont {Yi}},
  \bibinfo {author} {\bibfnamefont {K.-M.}\ \bibnamefont {Lee}}, \ and\
  \bibinfo {author} {\bibfnamefont {K.~I.}\ \bibnamefont {Goh}},\ }\href
  {http://arxiv.org/abs/1307.1253} {\enquote {\bibinfo {title} {Network
  robustness of correlated multiplex networks},}\ } (\bibinfo {year} {2013}),\
  \Eprint {http://arxiv.org/abs/1307.1253} {arXiv:1307.1253} \BibitemShut
  {NoStop}%
\bibitem [{\citenamefont {Valdez}\ \emph {et~al.}(2013)\citenamefont {Valdez},
  \citenamefont {Macri}, \citenamefont {Stanley},\ and\ \citenamefont
  {Braunstein}}]{valdez2013}%
  \BibitemOpen
  \bibfield  {author} {\bibinfo {author} {\bibfnamefont {L.~D.}\ \bibnamefont
  {Valdez}}, \bibinfo {author} {\bibfnamefont {P.~A.}\ \bibnamefont {Macri}},
  \bibinfo {author} {\bibfnamefont {H.~E.}\ \bibnamefont {Stanley}}, \ and\
  \bibinfo {author} {\bibfnamefont {L.~A.}\ \bibnamefont {Braunstein}},\ }\href
  {\doibase 10.1103/PhysRevE.88.050803} {\bibfield  {journal} {\bibinfo
  {journal} {Physical Review E}\ }\textbf {\bibinfo {volume} {88}},\ \bibinfo
  {pages} {050803(R)} (\bibinfo {year} {2013})},\ \Eprint
  {http://arxiv.org/abs/1308.4216} {arXiv:1308.4216} \BibitemShut {NoStop}%
\bibitem [{\citenamefont {Radicchi}\ and\ \citenamefont
  {Arenas}(2013)}]{Radicchi2013}%
  \BibitemOpen
  \bibfield  {author} {\bibinfo {author} {\bibfnamefont {F.}~\bibnamefont
  {Radicchi}}\ and\ \bibinfo {author} {\bibfnamefont {A.}~\bibnamefont
  {Arenas}},\ }\href@noop {} {\bibfield  {journal} {\bibinfo  {journal} {Nature
  Physics}\ }\textbf {\bibinfo {volume} {9}},\ \bibinfo {pages} {717} (\bibinfo
  {year} {2013})}\BibitemShut {NoStop}%
\bibitem [{\citenamefont {Chalupa}\ \emph {et~al.}(1979)\citenamefont
  {Chalupa}, \citenamefont {Leath},\ and\ \citenamefont {Reich}}]{chalupa1979}%
  \BibitemOpen
  \bibfield  {author} {\bibinfo {author} {\bibfnamefont {J.}~\bibnamefont
  {Chalupa}}, \bibinfo {author} {\bibfnamefont {P.~L.}\ \bibnamefont {Leath}},
  \ and\ \bibinfo {author} {\bibfnamefont {G.~R.}\ \bibnamefont {Reich}},\
  }\href {\doibase 10.1088/0022-3719/12/1/008} {\bibfield  {journal} {\bibinfo
  {journal} {J. Phys. C}\ }\textbf {\bibinfo {volume} {12}},\ \bibinfo {pages}
  {L31} (\bibinfo {year} {1979})}\BibitemShut {NoStop}%
\bibitem [{\citenamefont {Baxter}\ \emph {et~al.}(2010)\citenamefont {Baxter},
  \citenamefont {Dorogovtsev}, \citenamefont {Goltsev},\ and\ \citenamefont
  {Mendes}}]{BDG1}%
  \BibitemOpen
  \bibfield  {author} {\bibinfo {author} {\bibfnamefont {G.~J.}\ \bibnamefont
  {Baxter}}, \bibinfo {author} {\bibfnamefont {S.~N.}\ \bibnamefont
  {Dorogovtsev}}, \bibinfo {author} {\bibfnamefont {A.~V.}\ \bibnamefont
  {Goltsev}}, \ and\ \bibinfo {author} {\bibfnamefont {J.~F.~F.}\ \bibnamefont
  {Mendes}},\ }\href@noop {} {\bibfield  {journal} {\bibinfo  {journal} {Phys.
  Rev. E}\ }\textbf {\bibinfo {volume} {82}},\ \bibinfo {pages} {011103}
  (\bibinfo {year} {2010})}\BibitemShut {NoStop}%
\bibitem [{\citenamefont {Baxter}\ \emph {et~al.}(2011)\citenamefont {Baxter},
  \citenamefont {Dorogovtsev}, \citenamefont {Goltsev},\ and\ \citenamefont
  {Mendes}}]{baxter2011}%
  \BibitemOpen
  \bibfield  {author} {\bibinfo {author} {\bibfnamefont {G.~J.}\ \bibnamefont
  {Baxter}}, \bibinfo {author} {\bibfnamefont {S.~N.}\ \bibnamefont
  {Dorogovtsev}}, \bibinfo {author} {\bibfnamefont {A.~V.}\ \bibnamefont
  {Goltsev}}, \ and\ \bibinfo {author} {\bibfnamefont {J.~F.~F.}\ \bibnamefont
  {Mendes}},\ }\href {\doibase 10.1103/PhysRevE.83.051134} {\bibfield
  {journal} {\bibinfo  {journal} {Phys. Rev. E}\ }\textbf {\bibinfo {volume}
  {83}},\ \bibinfo {pages} {051134} (\bibinfo {year} {2011})}\BibitemShut
  {NoStop}%
\bibitem [{\citenamefont {Bollob{\'a}s}(2001)}]{bollobas2001}%
  \BibitemOpen
  \bibfield  {author} {\bibinfo {author} {\bibfnamefont {B.}~\bibnamefont
  {Bollob{\'a}s}},\ }\href@noop {} {\emph {\bibinfo {title} {Random graphs}}},\
  Vol.~\bibinfo {volume} {73}\ (\bibinfo  {publisher} {Cambridge university
  press},\ \bibinfo {year} {2001})\BibitemShut {NoStop}%
\bibitem [{\citenamefont {Dorogovtsev}\ \emph {et~al.}(2006)\citenamefont
  {Dorogovtsev}, \citenamefont {Goltsev},\ and\ \citenamefont
  {Mendes}}]{dorogovtsev2006}%
  \BibitemOpen
  \bibfield  {author} {\bibinfo {author} {\bibfnamefont {S.~N.}\ \bibnamefont
  {Dorogovtsev}}, \bibinfo {author} {\bibfnamefont {A.~V.}\ \bibnamefont
  {Goltsev}}, \ and\ \bibinfo {author} {\bibfnamefont {J.~F.~F.}\ \bibnamefont
  {Mendes}},\ }\href {\doibase 10.1103/PhysRevLett.96.040601} {\bibfield
  {journal} {\bibinfo  {journal} {Phys. Rev. Lett.}\ }\textbf {\bibinfo
  {volume} {96}},\ \bibinfo {pages} {040601} (\bibinfo {year}
  {2006})}\BibitemShut {NoStop}%
\bibitem [{\citenamefont {Cellai}\ \emph
  {et~al.}(2013{\natexlab{b}})\citenamefont {Cellai}, \citenamefont {Lawlor},
  \citenamefont {Dawson},\ and\ \citenamefont {Gleeson}}]{cellai2013b}%
  \BibitemOpen
  \bibfield  {author} {\bibinfo {author} {\bibfnamefont {D.}~\bibnamefont
  {Cellai}}, \bibinfo {author} {\bibfnamefont {A.}~\bibnamefont {Lawlor}},
  \bibinfo {author} {\bibfnamefont {K.~A.}\ \bibnamefont {Dawson}}, \ and\
  \bibinfo {author} {\bibfnamefont {J.~P.}\ \bibnamefont {Gleeson}},\ }\href
  {\doibase 10.1103/PhysRevE.87.022134} {\bibfield  {journal} {\bibinfo
  {journal} {Physical Review E}\ }\textbf {\bibinfo {volume} {87}},\ \bibinfo
  {pages} {022134+} (\bibinfo {year} {2013}{\natexlab{b}})},\ \Eprint
  {http://arxiv.org/abs/1209.2928} {arXiv:1209.2928} \BibitemShut {NoStop}%
\bibitem [{\citenamefont {Cellai}\ \emph {et~al.}(2011)\citenamefont {Cellai},
  \citenamefont {Lawlor}, \citenamefont {Dawson},\ and\ \citenamefont
  {Gleeson}}]{cellai2011}%
  \BibitemOpen
  \bibfield  {author} {\bibinfo {author} {\bibfnamefont {D.}~\bibnamefont
  {Cellai}}, \bibinfo {author} {\bibfnamefont {A.}~\bibnamefont {Lawlor}},
  \bibinfo {author} {\bibfnamefont {K.~A.}\ \bibnamefont {Dawson}}, \ and\
  \bibinfo {author} {\bibfnamefont {J.~P.}\ \bibnamefont {Gleeson}},\ }\href
  {\doibase 10.1103/PhysRevLett.107.175703} {\bibfield  {journal} {\bibinfo
  {journal} {Physical Review Letters}\ }\textbf {\bibinfo {volume} {107}},\
  \bibinfo {pages} {175703} (\bibinfo {year} {2011})}\BibitemShut {NoStop}%
\bibitem [{\citenamefont {Min}\ and\ \citenamefont {Goh}(2014)}]{Min2014}%
  \BibitemOpen
  \bibfield  {author} {\bibinfo {author} {\bibfnamefont {B.}~\bibnamefont
  {Min}}\ and\ \bibinfo {author} {\bibfnamefont {K.~I.}\ \bibnamefont {Goh}},\
  }\href {http://arxiv.org/abs/1401.1587} {\enquote {\bibinfo {title} {Multiple
  resource demands and viability in multiplex networks},}\ } (\bibinfo {year}
  {2014}),\ \Eprint {http://arxiv.org/abs/1401.1587} {arXiv:1401.1587}
  \BibitemShut {NoStop}%
\bibitem [{\citenamefont {Chen}\ and\ \citenamefont {Chao}(2011)}]{chen2011b}%
  \BibitemOpen
  \bibfield  {author} {\bibinfo {author} {\bibfnamefont {C.-Y.}\ \bibnamefont
  {Chen}}\ and\ \bibinfo {author} {\bibfnamefont {H.-C.}\ \bibnamefont
  {Chao}},\ }\href {\doibase 10.1002/sec.354} {\bibfield  {journal} {\bibinfo
  {journal} {Security and Communication Networks}\ } (\bibinfo {year} {2011}),\
  10.1002/sec.354}\BibitemShut {NoStop}%
\bibitem [{\citenamefont {Eschenauer}\ and\ \citenamefont
  {Gligor}(2002)}]{eschenauer2002}%
  \BibitemOpen
  \bibfield  {author} {\bibinfo {author} {\bibfnamefont {L.}~\bibnamefont
  {Eschenauer}}\ and\ \bibinfo {author} {\bibfnamefont {V.~D.}\ \bibnamefont
  {Gligor}},\ }in\ \href {\doibase 10.1145/586110.586117} {\emph {\bibinfo
  {booktitle} {Proceedings of the 9th ACM conference on Computer and
  communications security}}},\ \bibinfo {series and number} {CCS '02}\
  (\bibinfo  {publisher} {ACM},\ \bibinfo {address} {New York, NY, USA},\
  \bibinfo {year} {2002})\ pp.\ \bibinfo {pages} {41--47}\BibitemShut {NoStop}%
\bibitem [{\citenamefont {Chan}\ and\ \citenamefont {Perrig}(2005)}]{chen2005}%
  \BibitemOpen
  \bibfield  {author} {\bibinfo {author} {\bibfnamefont {H.}~\bibnamefont
  {Chan}}\ and\ \bibinfo {author} {\bibfnamefont {A.}~\bibnamefont {Perrig}},\
  }in\ \href {\doibase 10.1109/infcom.2005.1497920} {\emph {\bibinfo
  {booktitle} {INFOCOM 2005. 24th Annual Joint Conference of the IEEE Computer
  and Communications Societies. Proceedings IEEE}}},\ Vol.~\bibinfo {volume}
  {1}\ (\bibinfo  {publisher} {IEEE},\ \bibinfo {year} {2005})\ pp.\ \bibinfo
  {pages} {524--535 vol. 1}\BibitemShut {NoStop}%
\end{thebibliography}%

\end{document}